\documentclass[prd,preprintnumbers,twocolumn,eqsecnum,floatfix,a4paper,nofootinbib,superscriptaddress]{revtex4}
\usepackage{color}
\usepackage{calc}
\usepackage{amsmath,amssymb,graphicx}
\usepackage{amssymb,amsmath}
\usepackage{bm}
\usepackage{microtype}
\usepackage{booktabs}
\usepackage{times}
\usepackage[varg]{txfonts}
\usepackage[colorlinks, pdfborder={0 0 0}]{hyperref}
\usepackage[utf8]{inputenc}
\definecolor{LinkColor}{rgb}{0.75, 0, 0}
\definecolor{CiteColor}{rgb}{0, 0.5, 0.5}
\definecolor{UrlColor}{rgb}{0, 0, 0.75}
\hypersetup{linkcolor=LinkColor}
\hypersetup{citecolor=CiteColor}
\hypersetup{urlcolor=UrlColor}
\allowdisplaybreaks
\DeclareFontFamily{OT1}{pzc}{}
\DeclareFontShape{OT1}{pzc}{m}{it}{<-> s * [1.10] pzcmi7t}{}
\DeclareMathAlphabet{\mathpzc}{OT1}{pzc}{m}{it}

\newcommand{\io}{\iota}

\newcommand{\h}{\mathpzc{h}}

\newcommand{\B}{\mathpzc{B}}
\newcommand{\hlm}{\mathpzc{h}_{\ell m}}

\newcommand{\hc}{h_\times}
\newcommand{\hp}{h_+}

\newcommand{\fqnm}{f}
\newcommand{\sigmaqnm}{\sigma}

\begin{document}

\newcommand{\be}{\begin{equation}}
\newcommand{\ee}{\end{equation}}
\newcommand{\ber}{\begin{eqnarray}}
\newcommand{\eer}{\end{eqnarray}}
\def\bea{\begin{eqnarray}}
\def\eea{\end{eqnarray}}
\newcommand{\etal}{\emph{et al}}

\title{Accurate inspiral-merger-ringdown gravitational waveforms \\ for non-spinning black-hole binaries including the effect of subdominant modes}
\author{Ajit Kumar Mehta}
\affiliation{International Centre for Theoretical Sciences, Tata Institute of Fundamental Research, Bangalore 560012, India}
\author{Chandra Kant Mishra}
\affiliation{International Centre for Theoretical Sciences, Tata Institute of Fundamental Research, Bangalore 560012, India}
\affiliation{Indian Institute of Technology, Madras, Chennai 600036, India}
\author{Vijay Varma}
\affiliation{Theoretical Astrophysics, 350-17, California Institute of Technology, Pasadena, CA 91125, USA}
\affiliation{International Centre for Theoretical Sciences, Tata Institute of Fundamental Research, Bangalore 560012, India}
\author{Parameswaran~Ajith}
\affiliation{International Centre for Theoretical Sciences, Tata Institute of Fundamental Research, Bangalore 560012, India}

\begin{abstract}
We present an analytical waveform family describing gravitational waves (GWs) from the inspiral, merger and ringdown of non-spinning black-hole binaries including the effect of several non-quadrupole modes [($\ell = 2, m = \pm 1), (\ell = 3, m = \pm 3), (\ell = 4, m = \pm 4)$ apart from $(\ell = 2, m=\pm2)$]. We first construct spin-weighted spherical harmonics modes of \emph{hybrid} waveforms by matching numerical-relativity simulations (with mass ratio $1-10$) describing the late inspiral, merger and ringdown of the binary with post-Newtonian/effective-one-body waveforms describing the early inspiral. An analytical waveform family is constructed in frequency domain by modeling the Fourier transform of the hybrid waveforms making use of analytical functions inspired by perturbative calculations. The resulting highly accurate, ready-to-use waveforms are highly faithful (unfaithfulness $\simeq 10^{-4} - 10^{-2}$) for observation of GWs from non-spinning black hole binaries and are extremely inexpensive to generate. 
\end{abstract}
\preprint{LIGO-P1700160-v3}
\maketitle
\section{Introduction}
LIGO's recent observations of gravitational waves (GWs) from coalescing binary black hole systems~\cite{LSC_2016firstdetection, LSC_2016seconddetection, PhysRevLett.118.221101} mark the beginning of a new branch of astronomy. Based on the observed rate of GW signals, a large number of merger events can be expected in upcoming observing runs of Advanced LIGO and Virgo~\cite{LSC_2016rates, LSC_2016O1results}, providing us a unique opportunity to constrain the mass and spin distribution of binary black holes, to infer their astrophysical formation channels and to probe the true nature of extreme gravity. 

The most sensitive GW detection pipelines use the technique of matched filtering to detect GW signals from binary black holes~\cite{Usman:2015kfa,2017PhRvD..95d2001M}, which involves cross-correlating the data with theoretical templates of expected signals. Post detection, the physical and astrophysical properties of the GW source are inferred by comparing the data with theoretical signal templates, by means of Bayesian inference~\cite{Veitch:2014wba}. Tests of general relativity (GR) using GW observations also involves comparing the data with GR templates, to investigate the consistency of the observation with the prediction of GR~\cite{LSC_2016grtests}. Thus, accurate theoretical models of the expected signals are an essential input for GW astronomy. 

Theoretical templates describing the gravitational waveforms from the inspiral, merger and ringdown of binary black holes have been computed in the recent years by combining perturbative calculations in GR with large-scale numerical relativity simulations~\cite{Buonanno:1998gg,Buonanno:2000ef,Taracchini:2013rva,Bohe:2016gbl,Pan:2010hz,Pan:2011gk,Pan:2013rra,Damour:2009kr,Damour:2008gu,Damour:2012ky,Ajith:2007kx,Ajith:2009bn,Santamaria:2010yb,Hannam:2013oca,Khan_2016IMRPhenomD,Husa_2016IMRPhenomD}. Most of these waveform families aim to model only the leading (quadrupole; $\ell = 2, m = \pm 2$) modes of the gravitational radiation. Indeed, careful investigations suggested that the systematic errors introduced by neglecting subdominant (non-quadrupole) modes in the parameter estimation of the LIGO events are negligible~\cite{Abbott:2016wiq}. Due to the near ``face-on'' orientations of the binaries and moderate mass ratios, the effect of subdominant modes was negligible in the observed signals -- the systematic errors introduced by neglecting the subdominant modes were well within the statistical errors~\cite{Abbott:2016wiq}. However, for binaries with large mass ratios or high inclination angles or large signal-to-noise ratios, the systematic errors can dominate the statistical errors, biasing our inference of the physical and astrophysical properties of the source (see, e.g.,~\cite{Varma:2014hm, CalderonBustillo:2016hm, Varma:2016dnf}). In addition, including the effect of subdominant modes can improve the precision with which source parameters can be extracted, due to the increased information content in the templates (see, e.g.,~\cite{Sintes:1999cg,VanDenBroeck:2006ar,Arun:2007hu,Trias:2008pu,Arun:2008zn,Graff:2015bba,Lange:2017wki,OShaughnessy:2017tak}), potentially improving the accuracy of various observational tests of GR~\cite{Mishra:2010tp,Krishnendu:2017shb}.

In this paper we present an analytical waveform family describing GW signals from the inspiral, merger and ringdown of non-spinning black-hole binaries. These waveforms are constructed by combining perturbative calculations in GR with numerical-relativity (NR) waveforms in the ``phenomenological'' approach presented in a series of papers in the past~\cite{Ajith:2007qp,Ajith:2007kx,Ajith:2009bn,Santamaria:2010yb,Ajith:2007xh,Hannam:2013oca,Khan_2016IMRPhenomD,Husa_2016IMRPhenomD,London:2017aa}. This frequency domain, closed form waveform family has excellent agreement (faithfulness $> 0.99$) with ``target'' waveforms including subdominant modes, for binaries with mass ratio up to 10. Target waveforms including subdominant modes (with $\ell \leq 4, m \neq 0$) have been constructed by matching NR simulations describing the late inspiral, merger and ringdown of the binary with post-Newtonian (PN)/effective-one-body waveforms describing the early inspiral. Our highly accurate, ready-to-use, analytical waveforms are both effectual and faithful for observation of GWs from non-spinning black hole binaries and are extremely inexpensive to generate. 

This paper is organized as follows: Section~\ref{sec:methods} presents the construction of the analytical inspiral, merger, ringdown waveforms by combining numerical relativity with perturbative calculations in general relativity. In particular, Section~\ref{sec:hybrids} describes the construction of hybrid waveforms by matching the spherical harmonic modes of PN and NR waveforms, while Section~\ref{sec:phenom_model} describes the construction of the analytical waveform family approximating these hybrid waveforms in the frequency domain. The faithfulness of the new analytical waveforms to the original hybrids is studied in Section~\ref{sec:waveform_accuracy}. Section~\ref{sec:summary} presents some concluding remarks and discusses our future work. Supplementary calculations and information are presented in the Appendix. 

\section{The waveform model}
\label{sec:methods}

\begin{table}
\centering
\begin{tabular}{c@{\quad} c@{\quad}c@{\quad}c@{\quad}r}
\toprule
Simulation ID & $q$ & $M\omega_\mathrm{orb}$ & $e$ & \# orbits \\
\midrule
\emph{Fitting} \\ 
\midrule
SXS:BBH:0198 & $1.20$  &  $0.015$ &  $2.0 \times 10^{-4}$ &  $20.7$\\
SXS:BBH:0201 & $2.32$ &  $0.016$ &  $1.4 \times 10^{-4}$ &  $20.0$ \\
SXS:BBH:0200 & $3.27$ &  $0.017$ &  $4.1 \times 10^{-4}$ &  $20.1$ \\
SXS:BBH:0182 & $4.00$ &  $0.020$ &  $6.8 \times 10^{-5}$ &  $15.6$\\
SXS:BBH:0297 & $6.50$ & $0.021$ &  $5.9 \times 10^{-5}$ &  $19.7$\\
SXS:BBH:0063 & $8.00$ &  $0.019$ &  $2.8 \times 10^{-4}$ &  $25.8$ \\
SXS:BBH:0301 & $9.00$ &  $0.023$ &  $5.7 \times 10^{-5}$ &  $18.9$ \\
SXS:BBH:0185 & $9.99$ &  $0.021$ &  $2.9 \times 10^{-4}$ &  $24.9$ \\
\midrule
\emph{Verification}\\ 
\midrule
SXS:BBH:0066 & $1.00$ &  $0.012$ &  $6.4 \times 10^{-5}$ &  $28.1$\\
SXS:BBH:0184 & $2.00$ &  $0.018$ &  $7.6 \times 10^{-5}$ &  $15.6$\\
SXS:BBH:0183 & $3.00$ &  $0.019$ &  $6.3 \times 10^{-5}$ &  $15.6$\\
SXS:BBH:0182 & $4.00$ &  $0.020$ &  $6.8 \times 10^{-5}$ &  $15.6$\\
SXS:BBH:0187 & $5.04$ &  $0.019$ &  $5.0 \times 10^{-5}$ &  $19.2$\\
SXS:BBH:0181 & $6.00$ &  $0.017$ &  $7.9 \times 10^{-5}$ &  $26.5$\\
SXS:BBH:0298 & $7.00$ &  $0.021$ &  $4.0 \times 10^{-4}$ &  $19.7$\\
SXS:BBH:0063 & $8.00$ &  $0.019$ &  $2.8 \times 10^{-4}$ &  $25.8$ \\
SXS:BBH:0301 & $9.00$ &  $0.023$ &  $5.7 \times 10^{-5}$ &  $18.9$ \\
SXS:BBH:0185 & $9.99$ &  $0.021$ &  $2.9 \times 10^{-4}$ &  $24.9$ \\
\bottomrule
\end{tabular}
\caption{Summary of the parameters of the NR waveforms used in this paper: $q
\equiv m_1/m_2$ is the mass ratio of the binary, $M \omega_\mathrm{orb}$ is the
orbital frequency after the junk radiation and $e$ is the residual
eccentricity. The waveforms listed under the title \emph{Fitting} are used to 
produce the analytical fits described in Section~\ref{sec:phenom_model} while
those listed under the title \emph{Verification} are used for assessing the
faithfulness of the analytical model in Section~\ref{sec:waveform_accuracy}.}
\label{tab:NR_waveforms}
\end{table}

The two polarizations $h_+(t)$ and $h_\times(t)$ of GWs can be conveniently expressed as a complex waveform $\h(t) := h_+(t) - i \, h_\times(t)$. It is convenient to expand this in terms of the spin $-2$ weighted spherical harmonics so that the radiation along any direction $(\iota, \varphi_0)$ in the source frame can be expressed as 
\begin{equation}
\h(t; \iota, \varphi_0) = \sum_{\ell = 2}^{\infty} \sum_{m = -\ell}^{\ell} Y^{-2}_{\ell m}(\iota, \varphi_0) \, \hlm(t).
\end{equation}
The spherical harmonic modes $\hlm(t)$ are purely functions of the intrinsic parameters of the system (such as the masses and spins of the binary), while all the angular dependence is captured by the spherical harmonic basis functions $Y^{-2}_{\ell m}(\iota, \varphi_0)$. Here, by convention, the polar angle $\iota$ is measured with respect to the orbital angular momentum of the binary. The leading contribution to $\h(t; \iota, \varphi_0)$ comes from the quadrupolar ($\ell = 2, m=\pm 2$) modes. The relative contributions of various subdominant (nonquadrupole) modes depend on the symmetries of the system. For non-spinning binaries, it can be seen from the PN inspiral waveforms that the three subdominant modes with the largest amplitudes are $(\ell = 3, m = 3), (\ell = 4, m = 4)$ and $(\ell = 2, m = 1)$. This observation seems to hold through the merger regime (described by NR waveforms) as well. Thus, in this paper we focus on the modeling of these three subdominant modes, apart from the dominant quadrupole modes. Note that, due to the symmetry of non-spinning binaries, where the orbital motion is fully restricted to a fixed plane, the negative $m$ modes are related to positive $m$ modes by a complex conjugation. That is $h_{\ell -m} = (-)^\ell \, h_{\ell m}^*$. Also, the $m = 0$ modes are comprised of the nonlinear memory in the waveform, which has only negligible effect in GW detection and parameter estimation. It is also challenging to accurately extract this non-oscillatory signal from NR simulations~\cite{Favata:2010zu,Pollney:2010hs}. Thus, only $m > 0$ modes are considered in this paper. 
\subsection{Construction of hybrid waveforms}
\label{sec:hybrids}
In this paper we construct an analytical waveform family in the Fourier domain, that describes the three subdominant modes $(\ell m = 33, 44, 21)$ apart from the dominant $22$ mode of the GW polarizations from non-spinning black hole binaries. We start by constructing the spherical harmonic modes of \emph{hybrid} waveforms by combining PN and NR waveforms in a region where both calculations are believed to be accurate. 

PN inspiral waveforms, scaled to unit total mass and unit distance, can be written as   
\begin{equation}
\hlm^\mathrm{PN}(t) = 2 \, \eta \, v^2 \, \sqrt{\frac{16\pi}{5}} \, H_{\ell m} \, e^{-i\, m \, \varphi_\mathrm{orb}(t)},
\label{eq:pn_modes}
\end{equation}
where $\eta = m_1 m_2 / M^2$ is the symmetric mass ratio and $M = m_1+m_2$ is the total mass of the binary, $v = (M \omega_\mathrm{orb})^{1/3}$ is the PN expansion parameter, $ \omega_\mathrm{orb} = d\varphi_\mathrm{orb}/dt$ is the orbital frequency and $\varphi_\mathrm{orb}$ is the orbital phase. The PN mode amplitudes $H_{\ell m}$ are currently computed up to 3PN\footnote{The dominant, 22 mode inspiral model that we use here is actually 3.5PN accurate \cite{Faye:2012we}.} accuracy by~\cite{Blanchet:2008je,Kidder:2007rt,Arun:2004ff,Blanchet:1996pi} while the 3.5PN orbital phase $\varphi_\mathrm{orb}(t)$ can be computed in the adiabatic approximation using inputs given in \cite{Blanchet:2004ek} and references therein. 

In order to improve the accuracy of the inspiral waveforms, we compute the phase evolution of the inspiral part from the $22$ mode of the effective-one-body (EOB) waveforms calibrated to NR simulations (SEOBNRv4~\cite{Bohe:2016gbl}). Hence our inspiral waveforms are given by 
\begin{equation}
\hlm^\mathrm{PN}(t) = 2 \eta v^2 \sqrt{\frac{16\pi}{5}} \, H_{\ell m} \, e^{-i\, m \varphi_\mathrm{EOB22}(t)/2},
\label{eq:pn_modes_eobpn}
\end{equation}

where $\varphi_\mathrm{EOB22}$ is the phase of the $22$ mode of the
SEOBNRv4 waveform. Note that, for $m = 2$ modes, $H_{\ell m}$ contains imaginary
terms at order 2.5PN and above, which can be absorbed into the phase. However,
since this correction appears at order 5PN and above in the phase, we neglect
these corrections and use $|H_{\ell m}|$ instead of $H_{\ell m}$ for the $m =
2$ modes. 

Hybrid waveforms containing all the relevant modes ($\ell \leq 4, 1 \leq m \leq \ell$) 
are constructed by matching NR modes $\hlm^\mathrm{NR}(t)$ with PN modes $\hlm^\mathrm{PN}(t)$ 
with the same intrinsic binary parameters. The PN waveforms
are matched with NR by a least square fit over two rotations\footnote{These two
rotations are necessary due to the freedom in choosing the frame with respect
to which the NR and PN waveforms are decomposed into spherical harmonics modes.
In general three Euler rotations ($\iota, \varphi_0, \psi$) can be performed
between the two frames. However, one angle ($\iota$) is fixed by the choice of
aligning the $z$ axis along the direction of the total angular momentum of the
binary~\cite{Varma:2014hm,Bustillo:2015ova}.} on the NR waveform and the time-difference between NR and PN
waveforms over an appropriately chosen matching interval $(t_1,t_2)$, where
the NR and PN calculations are believed to be accurate.
\begin{equation} \mathrm{min}_{t_0,\varphi_0, \psi} \int_{t_1}^{t_2}
dt \sum_{\ell,m} \left|\hlm^\mathrm{NR}(t-t_0) \, e^{i (m \varphi_0 + \psi)}  -
\hlm^\mathrm{PN}(t) \, \right|.  \end{equation}
The hybrid waveforms are
constructed by combining the NR waveform with the ``best matched'' PN waveform
in the following way:
\begin{equation} \hlm^\mathrm{hyb}(t) \equiv \, \tau(t) \,
\hlm^\mathrm{NR}(t-t_0') \ e^{i(m\varphi_0'+\psi')} + [1-\tau(t)] \,
\hlm^\mathrm{PN}(t) , \end{equation}
where $t_0', \varphi_0'$ and $\psi'$ are the values of $t_0, \varphi_0$ and
$\psi$ that minimizes the difference $\delta$ between PN and NR waveforms.
Above, $\tau(t)$ is a weighting function defined by:
\begin{eqnarray} \tau(t) \equiv \left\{ \begin{array}{ll} 0 & \textrm{if $t <
t_1 $}\\ \frac{t-t_1}{t_2-t_1}  & \textrm{if $t_1 \leq t < t_2 $}\\ 1 &
\textrm{if $t_2 \leq t$.} \end{array} \right.  \label{eq:HybWaveWeight}
\end{eqnarray}
Our hybrid waveforms include spherical harmonic modes up to $\ell = 4$ and $m = -\ell ~
\mathrm{to} ~ \ell$ in this analysis, except the $m = 0$ modes. We use a subset of these
hybrid waveforms for constructing the analytical waveforms in the Fourier domain and 
to test the faithfulness of the analytical waveforms. The NR waveforms that were used to construct the hybrids are listed in Table~\ref{tab:NR_waveforms}. Note that, although the analytical waveforms only model the $22,33,44,21$ modes, their faithfulness is established by 
computing their mismatches with hybrids containing all the modes up to $\ell = 4$, except 
the $m=0$ modes.

\subsection{Construction of the analytical waveform model}
\label{sec:phenom_model}
%
\begin{figure*}[htb] \begin{center}
\includegraphics[width=5.5in]{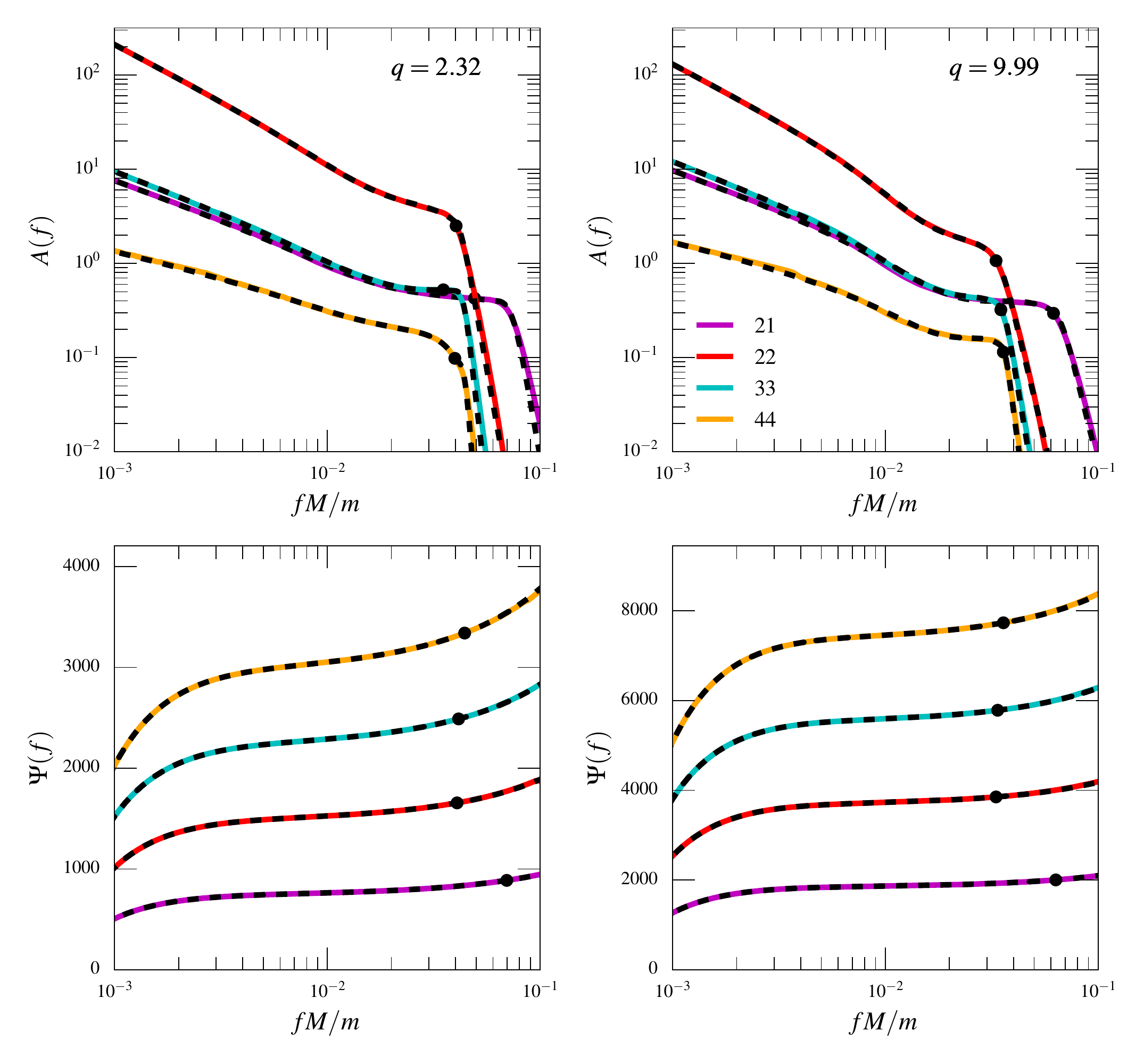}
\caption{Comparison between the amplitude (top panels) and phase (bottom panels) of the hybrids and analytical waveforms for selected mass ratios $q = 2.32$ (left panels) and $q = 9.99$ (right panels). In each plot, the solid lines correspond to hybrid waveforms for different modes and the dashed lines correspond to the analytical waveforms for the same mode. The legends show the $\ell m$ value for different modes. The black dots show the transition frequency  ($f^{\rm A}_{\ell m}$ and $f^{\rm P}_{\ell m}$) from the inspiral-merger to the ringdown part of the phenomenological amplitude and phase models.}
\label{fig:hybrid_phenom_comparison}
\end{center} \end{figure*}

In this section, we construct an analytical model for the Fourier transform $\hlm(f)$ of the real part of $\hlm(t)$ for the $22,33,44,21$ modes. Due to the symmetry of the non-spinning binaries, the Fourier transform of the imaginary part of $\hlm(t)$ can be computed by adding a phase shift of $\pi/2$ to $\hlm(f)$ (see Appendix \ref{sec:wf}). Writing this in terms of a Fourier domain amplitude and phase 
\begin{equation}
\hlm(f) = A_{\ell m}(f) \, e^{\mathrm{i} \, \Psi_{\ell m}(f)}, 
\end{equation}
our phenomenological model for the amplitude of each mode is the following: 
\begin{equation}
A_{\ell m}(f) = \left\{
\begin{array}{rl}
A_{\ell m}^\mathrm{IM}(f); \,\, f< f^{\rm A}_{\ell m}\\ \\  
A_{\ell m}^\mathrm{RD}(f); \,\, f \geq f^{\rm A}_{\ell m}.
\end{array} \right.
\label{eq:phenom_ampl_model}
\end{equation}
The Fourier frequencies below the matching frequency $f^{\rm A}_{\ell m}$ roughly correspond to the inspiral-merger stages of the signal, while the frequencies above $f^{\rm A}_{\ell m}$ roughly corresponds to the ringdown stage. The amplitude model for the inspiral-merger part is given by 
\begin{equation}
A_{\ell m}^\mathrm{IM}(f) = A_{\ell m}^\mathrm{PN}(f) \left(1+ \sum\limits_{k=0}^{k=1}\left(\alpha_{k,\,\ell m}\, +\alpha_{k,\,\ell m}^{L}\,\ln v_f \right)\,v_f^{k+8} \right),
\label{eq:phenom_ampl_model_IM}
\end{equation}
where $v_f = (2 \pi M f/m)^{1/3}$ and $A^\mathrm{PN}_{\ell m}(f)$ is the Pad\'e resummed version of the 3.5PN (3PN) amplitude of 22 (33, 44, 21)  mode in the Fourier domain (see Appendix \ref{sec:pade_amp}). The Pad\'e resummed version of the PN amplitude was employed to provide a better agreement with the late inspiral part of the hybrid amplitude. The inspiral-merger amplitude is modeled as the product of a Pad\`e resummed PN amplitude and another function that mimics a PN-like expansion. Such a form allows the resulting function to include very higher order terms, thus providing better fits to the late inspiral and merger part of the hybrid amplitude~\footnote{This idea is similar in spirit to the ``factorized resummed amplitude'' for effective one body waveforms proposed by~\cite{Damour:2008gu}.}. Above, $\alpha_{k,\,\ell m}$, $\alpha_{k,\,\ell m}^{L}$ and $f^{\rm A}_{\ell m}$ are phenomenological parameters whose values are determined from fits with numerical Fourier transforms of the hybrid waveforms.

The ringdown amplitude is modeled from the Fourier transform of a damped sinusoid, which is exponentially damped to mimic the high-frequency fall of the NR waveforms in the Fourier domain. That is, 
\begin{equation}
A_{\ell m}(f)^\mathrm{RD} =  w_{\ell m} ~ e^{-\lambda_{\ell m}} ~ \left|\B_{\ell m}(f) \right|,
\label{eq:phenom_ampl_model_RD}
\end{equation}
where $\B_{\ell m}(f)$ is the Fourier transform of the $\ell, m, n=0$ quasi-normal mode of a Kerr black hole with mass $M_f$ and dimensionless spin $a_f$~\cite{Berti:2009kk}, determined from initial masses:  
\begin{equation}
\B_{\ell m}(f) = \frac{\sigmaqnm_{\ell m} - i \, f}{\fqnm_{\ell m}^2 + (\sigmaqnm_{\ell m} - i \, f)^2}. 
\end{equation}
The frequencies  $\fqnm_{\ell m}$  and $\sigmaqnm_{\ell m}$ are the real and imaginary parts of the $\ell, m, n=0$ quasi-normal mode frequency $\Omega_{\ell m 0} = 2\pi \, (\fqnm_{\ell m} + \mathrm{i} \, \sigmaqnm_{\ell m})$. The phenomenological parameters $\lambda_{\ell m}$ in Eq.(\ref{eq:phenom_ampl_model_RD}) are determined from fits with numerical Fourier transforms of the hybrid waveforms, while $w_{\ell m}$ is a normalization constant to make the amplitudes continuous at the merger-ringdown matching frequency $f^\mathrm{A}_{\ell m}$. The mass $M_f$ and spin $a_f$ of the final black hole are computed from the masses $m_1$ and $m_2$ of the initial black holes, using fitting formulae calibrated to NR simulations. For this work, we use the fitting formulae given by~\cite{Pan:2011gk}. 

Our analytical model for the phase of the Fourier domain waveform reads 
\begin{equation}
\Psi_{\ell m}(f) = \left\{
\begin{array}{rl}
\Psi_{\ell m}^{\mathrm{IM}}(f) \quad\quad\quad ;\,\, f< f^{\rm P}_{\ell m}\\ \\
\Psi_{\ell m}^{\mathrm{RD}}(f) \quad\quad\quad ; \,\, f \geq f^{\rm P}_{\ell m}
\end{array} \right.\nonumber
\label{eq:phenom_phase_model}
\end{equation}
where the phase model for the inspiral-merger part of each mode takes the following form:
\begin{equation}
\Psi^{\rm IM}_{\ell m}(f) = \Psi^{\rm PN}_{\ell m} (f) + \sum_{k=0}^{k=4}(\beta_{k, \, \ell m}+\beta_{k, \, \ell m}^\mathrm{L}\,\ln v_f+ \beta_{k, \, \ell m}^\mathrm{L2}\,\ln^2 v_f)\,v_f^{k+8},
\label{eq:IMphase_model}
\end{equation}
where $\Psi^{\rm PN}_{\ell m} (f)$ is the PN phasing of the $\ell m$ mode, while the higher order phenomenological coefficients $\beta_{k, \, \ell m}$, $\beta_{k, \, \ell m}^\mathrm{L}$, $\beta_{k, \, \ell m}^\mathrm{L2}$ are determined from fits against the phase of hybrid waveforms. This particular phenomenological ansatz is motivated from the PN expansion of the frequency domain GW phasing of the inspiral waveforms in the test particle limit~(see, e.g., \cite{Varma:2013kna}).  

For the ringdown part of the phase we simply attach the phase of Fourier transform $\B_{\ell m}(f)$ of the $\ell, m, n=0$ quasi-normal mode at a transition frequency $f^\mathrm{P}_{\ell m}$. Thus, our ringdown phase model reads
\be
\label{eq:ringdown_phase_model}
\Psi^{\rm RD}_{\ell m}(f) = 2 \pi f t^\mathrm{P}_{\ell m} + \phi^\mathrm{P}_{\ell m} +  \arctan \, \B_{\ell m}(f),
\ee
where $t^\mathrm{P}_{\ell m}$ and $\phi^\mathrm{P}_{\ell m}$ are computed by matching two phases ($\Psi^{\rm IM}_{\ell m}$ and $\Psi^{\rm RD}_{\ell m}$) and their first derivative at the matching frequency $f^\mathrm{P}_{\ell m}$. Figure~\ref{fig:hybrid_phenom_comparison} provides a comparison of the amplitude and phase of the numerical Fourier transform of the hybrid waveforms, along with the analytical fits given by Eqs.~(\ref{eq:phenom_ampl_model}) and (\ref{eq:phenom_phase_model}).

\begin{figure*}[htb] 
\includegraphics[width=6.3in]{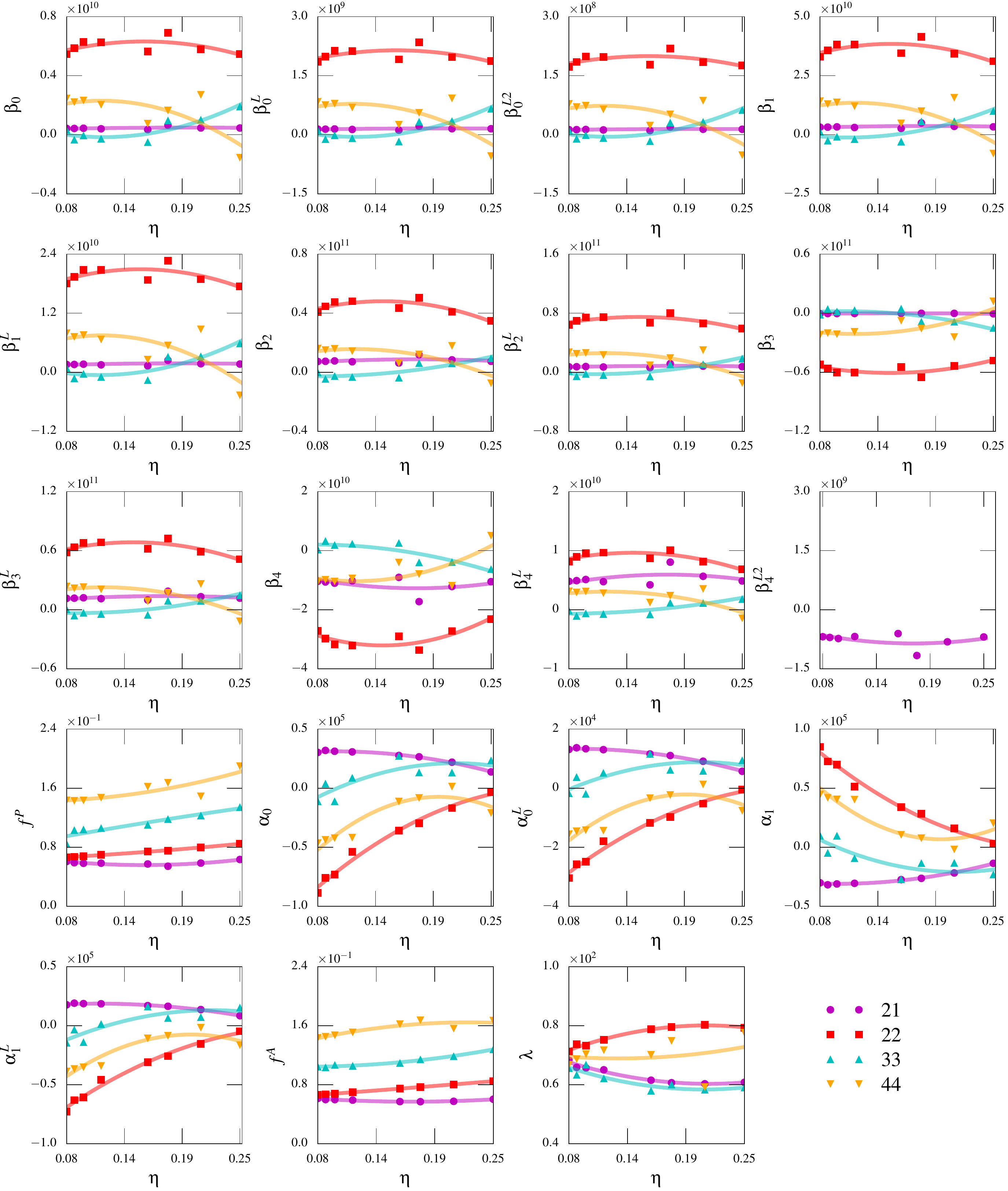}
\vspace{4mm}
\caption{The estimated values of the phenomenological parameters describing the analytical waveforms, plotted against the symmetric mass ratio $\eta$. Different markers correspond to different modes. Also plotted are the fits given by Eqs.~(\ref{eq:phenfits_phase_fits}).}
\label{fig:phenpars_vs_eta}
\end{figure*}

Finally, the phenomenological parameters describing the analytical model are represented as quadratic functions of the symmetric mass ratio $\eta$
\begin{eqnarray}
\label{eq:phenfits_phase_fits}
\alpha_{i, \, \ell m} &=& a_{i, \, \ell m} + b_{i, \, \ell m}\,\eta + c_{i, \, \ell m}\,\eta^2\,, \nonumber \\
\alpha_{i, \, \ell m}^{\rm L} &=& a_{i, \, \ell m}^{\rm L} + b_{i, \, \ell m}^{\rm L} \,\eta + c_{i, \, \ell m}^{\rm L}\,\eta^2\,, \nonumber \\
\beta_{k, \, \ell m} &=& a_{k, \, \ell m} + b_{k, \, \ell m}\,\eta + c_{k, \, \ell m}\,\eta^2\,, \nonumber \\
\beta_{k, \, \ell m}^{\rm L} &=& a_{k, \, \ell m}^{\rm L} + b_{k, \, \ell m}^{\rm L} \,\eta + c_{k, \, \ell m}^{\rm L}\,\eta^2\,, \nonumber \\
\beta_{j, \, \ell m}^{\rm L2} &=& a_{j, \, \ell m}^{\rm L2} + b_{j, \, \ell m}^{\rm L2} \,\eta + c_{j, \, \ell m}^{\rm L2}\,\eta^2\,, \nonumber \\
f^\mathrm{A}_{\ell m}  &=&  ( a_{\ell m}^{\rm L} + b_{\ell m}^{\rm L} \,\eta + c_{\ell m}^{\rm L}\,\eta^2)\, / M, \nonumber \\
f^\mathrm{P}_{\ell m}  &=&  ( a_{\ell m}^{\rm L} + b_{\ell m}^{\rm L} \,\eta + c_{\ell m}^{\rm L}\,\eta^2)\, / M, 
\end{eqnarray}
where the index $i$ runs from 0 to 1,  $k$ runs from 0 to 4 and $j$ is 0 except for $21$ mode ($j$=0,1). Figure~\ref{fig:phenpars_vs_eta} shows the values of the phenomenological parameters estimated from the hybrid waveforms, as well as the fits described by Eq.~(\ref{eq:phenfits_phase_fits}). 

%

\subsection{Assessing the accuracy of the analytical model}
\label{sec:waveform_accuracy}

\begin{figure*}[htb] \begin{center}
\includegraphics[width=3.4in]{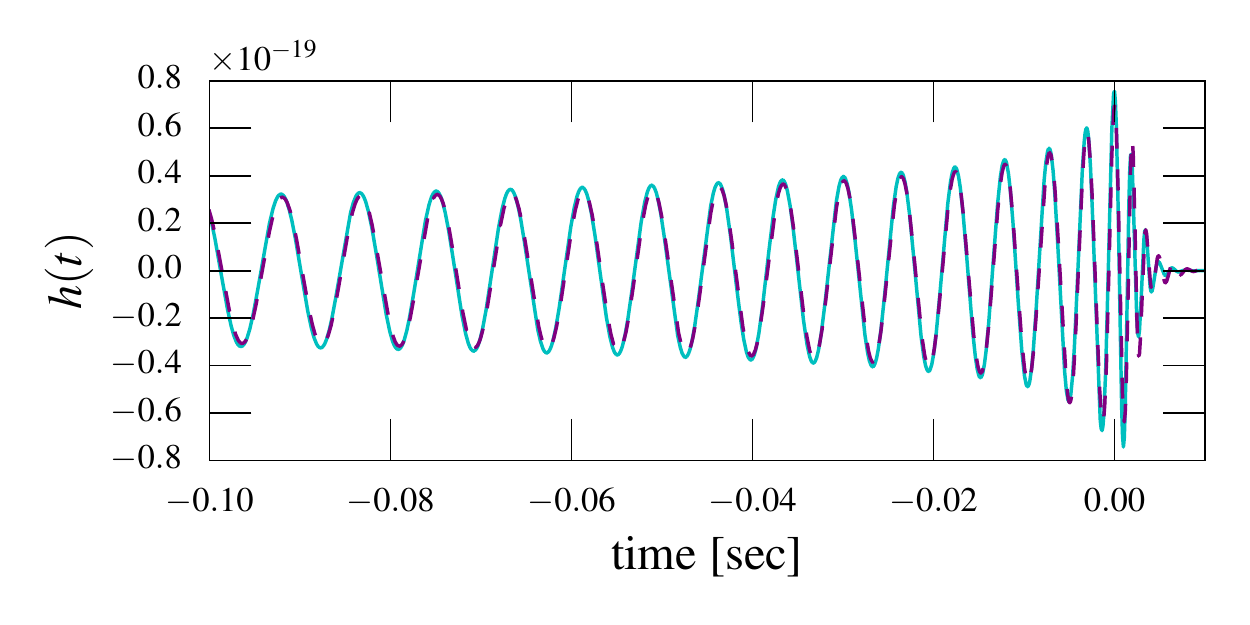}
\includegraphics[width=3.4in]{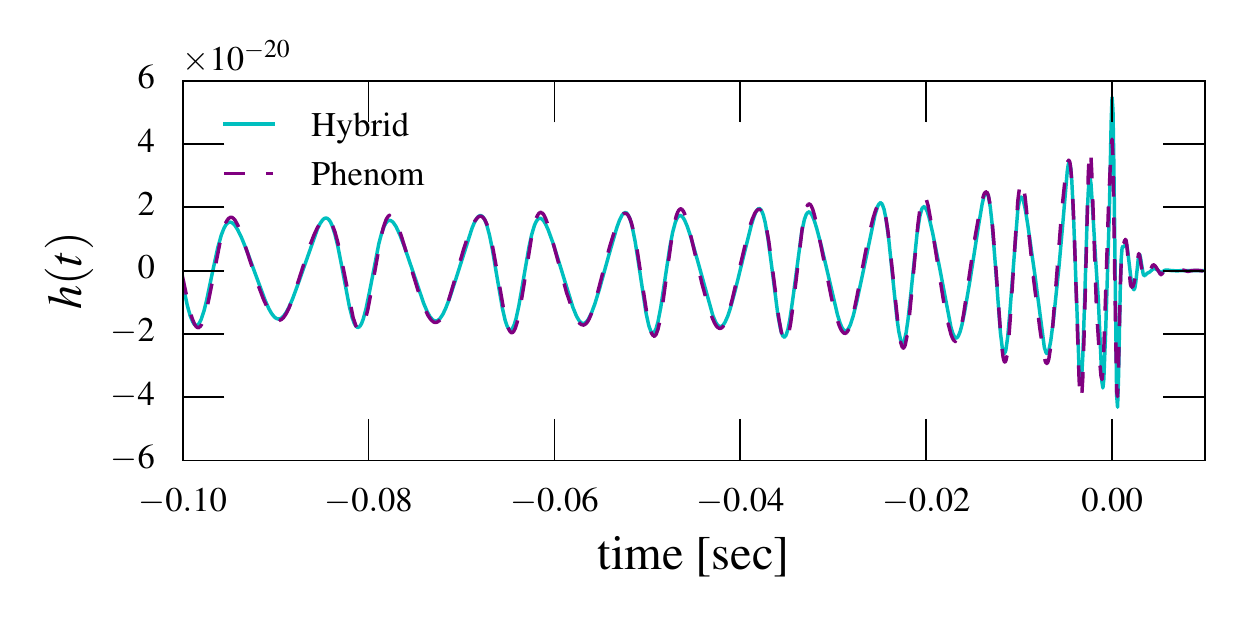}
\caption{Comparison between hybrid waveforms and our analytical phenomenological waveforms for a binary with total mass $M = 20 M_\odot$ and mass ratio $q = 10$. Hybrid waveforms are constructed using all the modes with $\ell \leq 4$, except the $m = 0$ modes. Phenomenological waveforms are constructed by taking the (discrete) inverse Fourier transform of the analytical waveforms in the Fourier domain. The left panel corresponds to a ``face-on'' binary (inclination angle $i=0.00$) while the right panel corresponds to an ``edge-on'' binary ($i=1.57$).}
\label{fig:hybrid_phenom_comparison_td}
\end{center} \end{figure*}

\begin{figure*}[htb] \begin{center}
\includegraphics[width=7.1in]{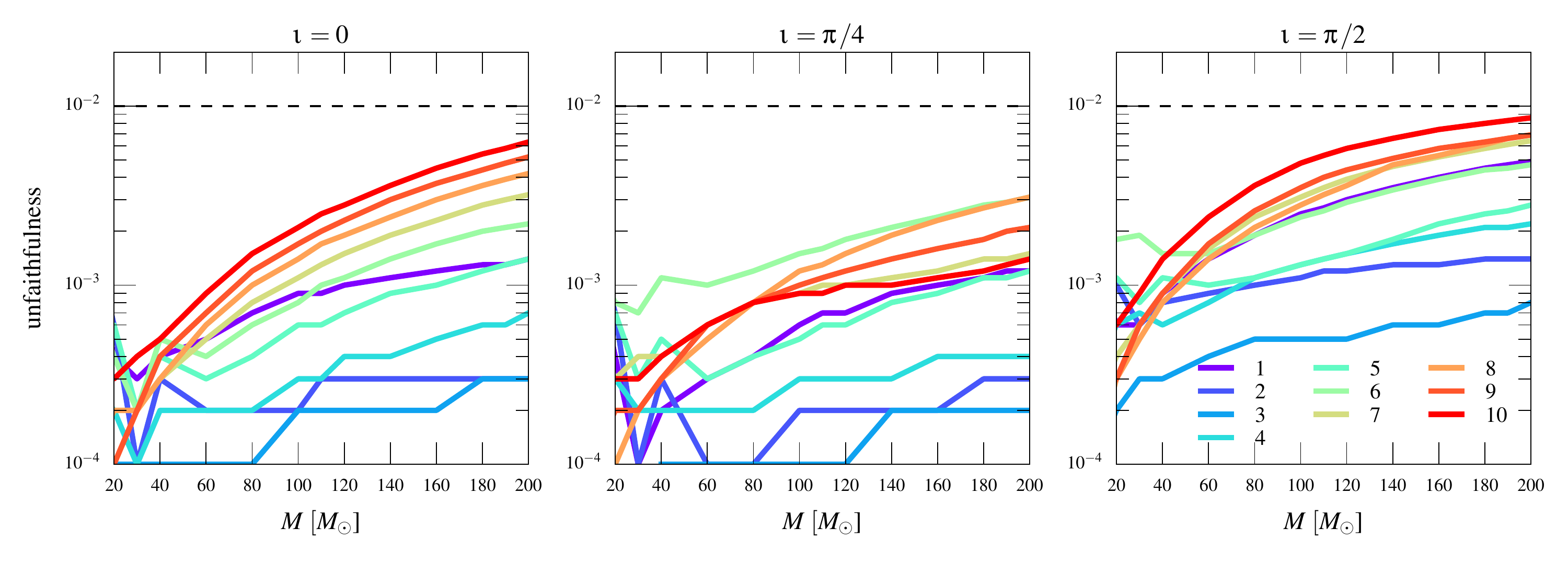}
\vspace*{-2mm}
\caption{The unfaithfulness (mismatch) of the analytical waveform family towards hybrids for various inclination angles $\iota$. The horizontal axes report the total mass of the binary and different curves correspond to different mass ratios $q$ (shown in the legend). Horizontal black dashed lines correspond to a mismatch of 1\%. The overlaps are computed assuming the design power spectrum of Advanced LIGO (in the ``high-power, zero-detuning'' configuration~\cite{adligo-psd}), assuming a low-frequency cutoff of 20 Hz.} 
\label{fig:mismatch_iota}
\end{center} \end{figure*}

Here we quantify the faithfulness of the analytical model that we constructed by computing the mismatches of these with the hybrid waveforms, which are assumed as our fiducial waveforms. Indeed, relative contribution of different modes depend on the orientation of the binary with respect to the line of sight. Figure~\ref{fig:hybrid_phenom_comparison_td} shows some examples of the hybrid waveforms for different orientations along with the corresponding waveforms generated from our analytical model (by taking the inverse Fourier transform). Computation of these polarizations $h_+(t)$ and $h_\times(t)$ is described in Appendix~\ref{sec:wf}. Polarizations of the hybrid waveforms have been computed using all the modes up to $\ell = 4$, except the $m = 0$ modes; see, Eq.~(\ref{eq:complexh}), while the analytical phenomenological waveforms have been computed using $\ell m = 22, 21, 33, 44$ modes only, by computing the inverse Fourier transform of the expression Eq.~(\ref{eq:hphc_freqdomain_final}) numerically.

Since the relative contribution to the observed $h(t)$ from different modes depend on the relative orientation of the binary, the mismatches of our analytical phenomenological waveforms with the hybrids will be a function of the orientation angles. Figure~\ref{fig:mismatch_iota} shows the mismatches for different orientations as a function of the total mass $M$ and mass ratio $q$ of the binary. Relative contribution from subdominant modes are expected to be the smallest [largest] for binaries with inclination angles $\iota = 0$ [$\iota = \pi/2$]. The figure shows that the mismatches are less than 1\% for all orientations, illustrating the high faithfulness of our phenomenological waveforms. Note that GW detectors have a strong selection bias towards small inclination angles ($\iota \rightarrow 0$). Hence, the mismatches averaged over all orientations are likely to be comparable to the ones reported in the left panel of the Figure ($\iota = 0$).  

\section{Summary and conclusions}
\label{sec:summary}
In this paper, we presented an analytical family of frequency-domain waveforms describing the GW signals from non-spinning black-hole binaries, including some of the leading subdominant modes of the radiation $(\ell m = 21,33,44)$, apart from the dominant ($\ell m = 22$) mode. The construction of these analytical waveforms involves two major steps: 1) the construction of a set of hybrid waveforms by combining the spherical harmonic modes of PN and NR waveforms corresponding to a limited set of mass ratios $1 \leq q \leq 10$, 2) representing the numerical Fourier transform of the hybrid waveforms by a suitable set of analytical functions which allow us to interpolate these waveforms smoothly over the parameter space. The analytical gravitational waveforms that are constructed in this way are highly faithful (mismatch $0.01\% - 1\%$) to our target hybrid waveforms that include all the modes up to $\ell = 4$ (except the $m = 0$ modes).  

The Fourier domain amplitude of our phenomenological waveforms contain a inspiral-merger part that is smoothly matched to the ringdown part. The inspiral-merger amplitude is modeled as the product of a Pad\`e resummed version of the Fourier domain PN amplitude and another function that mimics a PN-like expansion whose coefficients are determined by fitting against the Fourier-domain amplitude of the hybrid waveforms. The ringdown part is modeled as the Fourier transform of a time-symmetric damped sinusoid, which is exponentially damped to mimic the high-frequency fall of the NR waveforms in Fourier domain. Similarly, the Fourier domain phase is modeled as a PN-like series including the known coefficients from PN theory till 3.5PN order, while the higher order ``pseudo-PN'' terms are determined by fitting against the hybrid waveforms. The resulting waveforms are also computationally inexpensive to generate, allowing their direct implementation in GW searches and parameter estimation.

A note on the limitations of this work: These waveforms aim to model the GW signals from non-spinning black hole binaries in quasicircular orbits. Spin effects of black holes are not considered. (We note that, an approximate phenomenological model for spinning binaries, making use of rescaled amplitudes and frequencies of the $\ell = m = 2$ mode for modeling the non-quadrupole modes has been developed recently~\cite{London:2017aa} and an EOB model is under development~\cite{Cotesta:2017xxx}). Additionally, we consider only a subset of the subdominant modes $\ell m = 21, 33, 44$. Although the subdominant modes that we neglect here makes no  appreciable contributions to the total signal for the mass ratios that we consider, this may not be the case for even higher mass ratios. Modeling of some of the subdominant modes (e.g., $\ell m = 32, 43$, etc.) that we neglect here could be harder, due to the effect of ``mode-mixing''~\cite{Berti:2014fga}. There is ongoing work that aims to include the spin effects, to model the subdominant modes that are neglected here, and to extend the validity of these waveforms making use of numerical waveforms modeling binary black holes with extreme mass ratios. 

\acknowledgments 
We are indebted to the SXS collaboration for making a public catalog of numerical-relativity waveforms. We thank K. G. Arun, Bala Iyer, Sascha Husa, Mark Hannam and Nathan Johnson-McDaniel for very useful discussions. P.~A, A.~K.~M and V.~V acknowledge support from the Indo-US Centre for the Exploration of Extreme Gravity funded by the Indo-US Science and Technology Forum (IUSSTF/JC-029/2016). In addition, P.~A.'s research was supported by the AIRBUS Group Corporate Foundation through a chair in ``Mathematics of Complex Systems'' at the International Centre for Theoretical Sciences (ICTS), by a Ramanujan Fellowship from the Science and Engineering Research Board (SERB), India, by the SERB FastTrack fellowship SR/FTP/PS-191/2012, and by the Max Planck Society through a Max Planck Partner Group at ICTS. V.~V.'s research was supported by NSF Grant PHY-1404569 to Caltech and the Sherman Fairchild Foundation. Computations were performed at the ICTS clusters Mowgli, Dogmatix, and Alice. This document has LIGO preprint number LIGO-P1700160-v3. 

\appendix

\begin{figure*}[htb] 
\begin{center}
\includegraphics[width=5.5in]{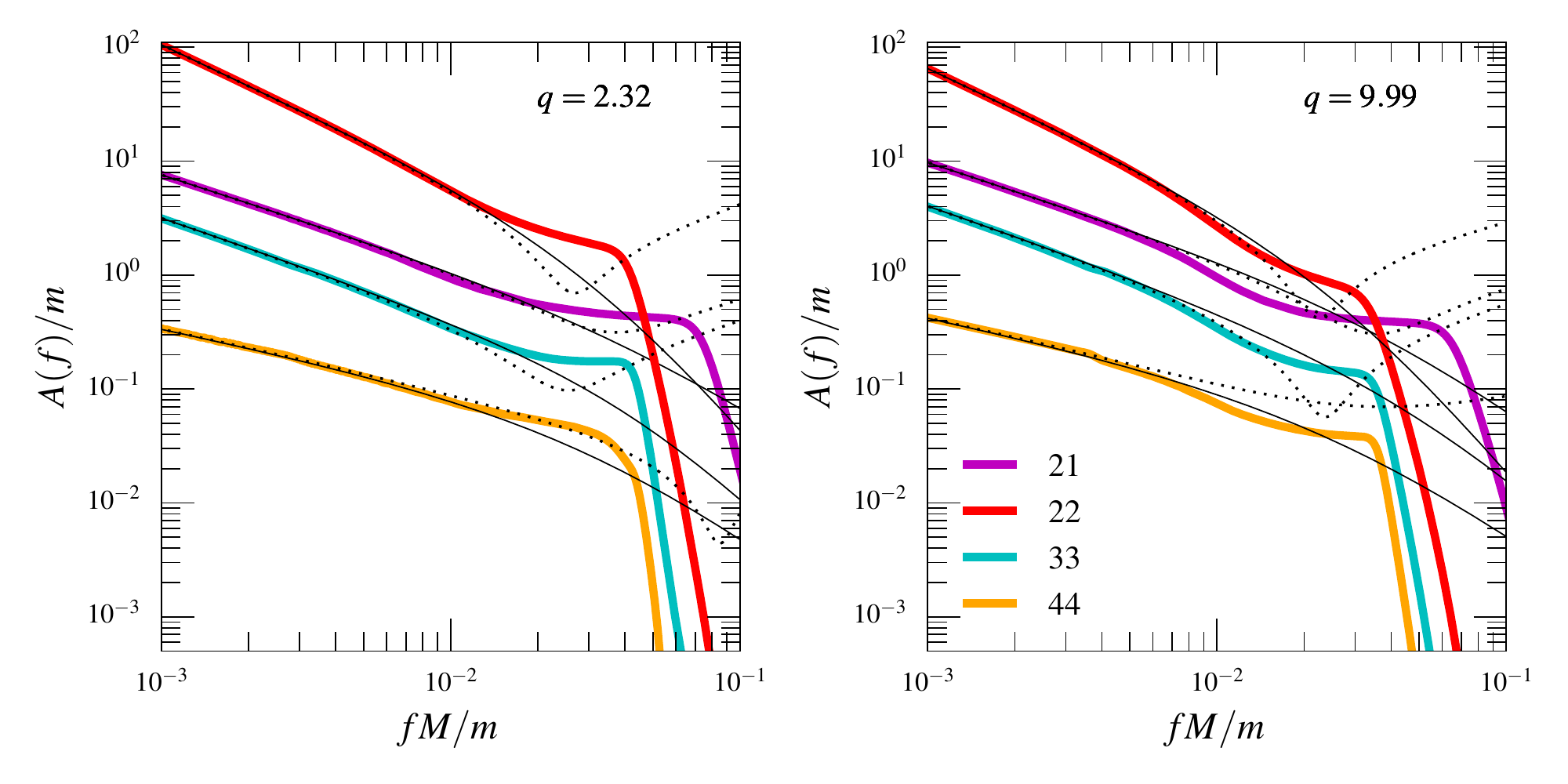}
\caption{Comparison of the Pad\'e approximant of the PN inspiral amplitude (thin, solid lines) with regular Taylor expanded amplitude (thin, dotted lines) and the amplitude of the hybrid waveform (thick, solid lines) for different modes $\ell m = 21, 22, 33, 44$. The left panel corresponds to mass ratio $q = 2.32$ while the right panel corresponds to mass ratio $q \simeq 10$.} 
\label{fig:pade_taylor_ampl_comparison}
\end{center} 
\end{figure*}

\section{Pad\'e summation on the post-Newtonian amplitude in the Fourier domain}
\label{sec:pade_amp}
The PN expression for various mode amplitudes have a point of inflection at high frequencies when higher order PN corrections are included. This makes it inconvenient to model the amplitude of the phenomenological waveforms as a factorized correction to the PN waveforms as shown in Eq.(\ref{eq:phenom_ampl_model_IM}). In order to resolve this issue, and to generally improve the agreement of PN amplitude with that of the hybrid waveforms, we construct our inspiral amplitude model by performing Pad\'e summation of these expressions. Pad\'e summation of a given function involves finding a suitable rational function whose Taylor expansion to a given order matches exactly with the Taylor expression of the original function to the same order. For instance, Pad\'e summation of a simple power series $\sum_{k=0}^{n} a_n x^n$ can be written as 
\be
\label{eq:pade_example}
P_{q}^{p}(x) =\frac{\sum_{k=0}^{p} b_k x^k}{\sum_{k=0}^{q} c_k x^k}\,,
\ee
where $p+q=n$. Each of these coefficients ($b_k$ and $c_k$) then can readily be obtained by demanding that a Taylor expansion of the above to order $n$ reproduces exactly the first $n$ terms the given power series. Such rational functions are called Pad\'e approximants (see App.~A of Ref.~\cite{DIS98} for a related discussion). 

After comparing various Pad\'e approximants corresponding to PN amplitude expressions for each mode we find the most suitable (i.e., an approximant with no point of inflection) approximant corresponds to the choice of rational functions associated with $p=0$ and $q=n$, i.e., $P^{0}_{n}$. For instance for the $\ell=m=2$ mode whose (normalized) amplitude is given by the series, $\sum_{k=0}^{7}\alpha_k v^k$, the Pad\'e approximant we find most suitable for our purposes is given by 
\be
\label{eq:pade22_example}
P_{7}^{0}(v) =\frac{\beta_0}{\sum_{k=0}^{7} \gamma_k v^k}\,.
\ee
Moreover by the virtue of the use of normalized amplitude expressions in constructing the Pad\'e approximants we can choose (without any loss of generality), $\beta_0=\gamma_0=1$, which leads to the following simple expression
\be
\label{eq:pade22_example2}
P_{7}^{0}(v) =\frac{1}{1+\sum_{k=1}^{7} \gamma_k v^k}\,.
\ee
Figure~\ref{fig:pade_taylor_ampl_comparison} shows a comparison of the standard Taylor expanded 3.5PN (3PN) amplitude for 22 (21, 33, 44) with our corresponding resummed Pad\'e function as well as the amplitude of the hybrid waveform in the Fourier domain. Explicit expressions for Pad\'e approximants for modes we consider here are listed in Appendix \ref{sec:fd_wf_pade} below.

\section{Pad\'e resummed frequency domain expressions for the inspiral amplitude}
\label{sec:fd_wf_pade}

As discussed above, occurrence of divergences in the PN amplitudes when including 
higher PN terms motivates us to find Pad\'e resummed expressions of the PN amplitudes 
as our inspiral amplitude model. Here we provide, analytical expression for the complete 
inspiral model for each mode in the frequency domain which are constructed using the 
prescription listed in Ref.~\cite{VanDenBroeck:2006qu} and uses Stationary Phase 
Approximation. Resulting expression for each mode of the gravitational wave
polarizations in the frequency domain take the following general form

\begin{equation} \tilde{h}_{\ell m}(f) = \frac{M^2}{D_L}\,\pi\,
\sqrt{\frac{2 \eta}{3}}\;v_f^{-7/2}\;e^{-{\rm i} \, m\,\Psi(v_f)}
\;{H_{\ell m}}(v_f) \,.
\end{equation}

Here, $M$ and $\eta$ again denote the total mass and symmetric mass ratio parameter of the binary whereas $D_L$ is the luminosity distance of the source. The quantity $v_f$ is given by $v_f \equiv (2\,\pi M f/m)^{1/3}$ and $\Psi(v_f)$ represents the orbital phase of the binary computed using stationary phase approximation (see for instance Ref.~\cite{VanDenBroeck:2006qu} for a related discussion). Finally, $H_{\ell m}$ are the Pad\'e resummed version of the inspiral amplitudes and takes following form for the modes whose complete models are presented in this study. They read

\begin{subequations}
\begin{align}
H_{22} &= {}^{22}P_{7}^{0}(v_f)\\
H_{21} &= \mathrm{i} \, \frac{{\sqrt2 }}{3}\,\delta\, [{}^{21}P_{5}^{0}(v_f)]\,v_f\\
H_{33} &= -\mathrm{i} \, {3\over 4}\sqrt {5\over7}\delta\,[{}^{33}P_{5}^{0}(v_f)]\,v_f\\
H_{44} &= - {4\over 9}\sqrt {10\over7} (1-3\,\eta)\, [{}^{44}P_{4}^{0}(v_f)]\,v_f^2
\end{align}
\label{eq:Hlm}
\end{subequations}
here, ${}^{\ell m}P_{n}^{0}(v_f)$ are (normalized) amplitude expressions for Pad\'e resummed expressions for inspiral amplitudes corresponding to $p=0$ and $q=n$ (see Appendix \ref{sec:pade_amp} for related discussions) and can be expressed in the following general form.
\be
\label{eq:pade0n_mode}
{}^{\ell m}P_{n}^{0}(v) =\frac{1}{1+\sum_{k=1}^{n} \gamma^{\ell m}_k v^k}\,,
\ee
where $\gamma^{\ell m}$ corresponding to each mode can be written in the following form,
\begin{subequations}
\begin{align}
\gamma_1^{22} &= 0\\
\gamma_2^{22} &= \frac{323}{224}-\frac{451}{168} \eta\\
\gamma_3^{22} &= 0\\
\gamma_4^{22} &= \frac{44213383}{8128512}-\frac{92437}{48384} \eta +\frac{483509}{169344} \eta ^2\\
\gamma_5^{22} &= \frac{85 \pi }{64}+\left(24 \,{\rm i}-\frac{85 \pi }{16}\right) \eta\\
\gamma_6^{22} &=\frac{40919017211}{1226244096}-\frac{428 \,{\rm i} \,\pi }{105}+\left(-\frac{1906061676931}{15021490176}\right.\nonumber\\&+\left. \frac{205 \,\pi ^2}{48}\right) \eta+\frac{6864704395}{1251790848} \eta ^2-\frac{48013667}{34771968} \eta ^3\\
\gamma_7^{22} &= \frac{633281 \pi }{1161216}+\left(\frac{2357 \,{\rm i}}{324}-\frac{21367 \,\pi }{3456}\right) \eta +\left(-\frac{86519 \,{\rm i}}{945}\right.\nonumber\\&+\left.\frac{496409 \,\pi }{24192}\right) \eta ^2\\
\label{eq:fd_22amp_pade}
\gamma_1^{21} &= 0\\
\gamma_2^{21} &= -\frac{335}{672}-\frac{117}{56} \eta \\
\gamma_3^{21} &= \frac{\rm i}{2}+\pi +2\,{\rm i} \ln 2\\
\gamma_4^{21} &= \frac{2984407}{8128512}+\frac{62659}{12544} \eta +\frac{96847}{56448} \eta ^2\\
\gamma_5^{21} &= -\frac{335 {\rm i}}{1344}+\frac{1115 \pi }{1344}+\eta  \left(\frac{1255 {\rm i}}{112}-\frac{885 \pi }{112}-\frac{145}{28} {\rm i} \ln 2\right)\nonumber\\&-\frac{335}{336} {\rm i} \ln 2\\
\label{eq:fd_21amp_pade}
\gamma_1^{33} &= 0\\
\gamma_2^{33} &= \frac{1945}{672}-\frac{27}{8} \eta \\
\gamma_3^{33} &= \frac{2\,{\rm i}}{5}-\pi +6\,{\rm i} \ln 2-6\,{\rm i} \ln 3\\
\gamma_4^{33} &=\frac{4822859617}{447068160}-\frac{5571877}{887040} \eta +\frac{301321}{63360} \eta ^2\\
\gamma_5^{33} &= \frac{389\,{\rm i}}{32}-\frac{2105 \pi }{1344}-\frac{1945\, {\rm i}}{112} \ln ({3/2})+\eta  \left(\frac{33079\,{\rm i}}{1944}\right.\nonumber\\&-\left.\frac{23 \pi }{16}+\frac{93\,{\rm i} }{4} \ln (3/2)\right)\\
\label{eq:fd_33amp_pade}
\gamma_1^{44} &= 0\\
\gamma_2^{44} &= \frac{1}{1-3 \eta} \left(-\frac{158383}{36960}+\frac{128221}{7392} \eta -\frac{1063}{88} \eta ^2\right)\\
\gamma_3^{44} &=\frac{1}{1-3 \eta} \left(-\frac{42\,{\rm i}}{5}+2 \pi +\eta  \left(\frac{1193\,{\rm i}}{40}-6 \pi -24 {\rm i} \ln 2\right)
\right.\nonumber\\&+\left.8 \,{\rm i} \ln 2\right)\\
\gamma_4^{44} &= \frac{1}{(1-3 \eta)^2} \left(\frac{5783159561419}{319653734400}-\frac{6510652977943}{53275622400} \eta \right.\nonumber\\&+\left.\frac{8854729392203}{35517081600} \eta^2-\frac{1326276157}{8456448} \eta ^3+\frac{63224063 }{1006720}\eta ^4\right).
\label{eq:fd_44amp_pade}
\end{align}
\end{subequations}

Finally, the orbital phase takes the following form in Fourier domain 

\begin{equation} \Psi(v_f) = 2\pi f t_0 - \pi/4 + \frac{3}{256 \, \eta\,v_f^{5}} \left[\sum_{k=0}^{7} \psi_{k} v_f^k\right],  
\end{equation}
where, $t_0$ represents a reference time\footnote{Note that we have set the phase at reference time to zero, since phase shifts can be introduced on the waveform by the spherical harmonic basis functions; see Eq.(\ref{eq:hphc_freqdomain_final}).} and $\psi_k$ denote the PN corrections to the leading order orbital phase. These read
\begin{subequations}
\begin{align}
\psi_0 & = 1,\\
\psi_1 & = 0,\\
\psi_2 & = \frac{3715}{756}+\frac{55}{9} \eta, \\
\psi_3 & =-16 \, \pi,\\
\psi_4 & = \frac{15293365}{508032}+\frac{27145 }{504} \eta+\frac{3085 }{72}\eta ^2,\\
\psi_5 & =\pi \left(\frac{38645}{756}-\frac{65}{9}\eta\right)(1+ 3 \ln v_f), \\
\psi_6 & = \frac{11583231236531}{4694215680}-\frac{6848 \gamma_{\rm E}}{21}-\frac{640 \pi ^2}{3}\nonumber\\&+\left(-\frac{15737765635}{3048192}+\frac{2255 \pi ^2}{12}\right) \eta +\frac{76055}{1728} \eta^2-\frac{127825}{1296}\eta^3\nonumber\\&-\frac{6848}{21} \ln (4 v_f),\\
\psi_7 &=\frac{77096675 \pi}{254016}+\frac{378515 \pi}{1512} \eta -\frac{74045 \pi }{756} \eta^2, 
\label{eq:fd_phase}
\end{align}
\end{subequations}
where $\gamma_{\rm E}$ is the Euler's constant. 

\section{Computing the $+$ and $\times$ polarization waveforms from the spherical harmonic modes in the frequency domain}
\label{sec:wf}

The complex time-series, $\h=\hp-i\,\hc$, can be decomposed into a sum of spherical harmonic modes as

\begin{equation} 
\label{eq:complexh} 
\h (t) =\sum^{+\infty}_{\ell=2}\sum^{\ell}_{m=-\ell} \hlm(t)\,Y^{\ell
m}_{-2}(\io,\varphi_0)\,, 
\end{equation}
where $Y^{\ell m}_{-2}$'s (the spin-weighted spherical harmonics of weight
$-2$) are functions of the spherical angles $(\io,\varphi_0)$ defining the
binary's orientation, and are given as
\begin{equation}
Y^{\ell m}_{-2} = \sqrt{\frac{2\ell+1}{4\pi}}\,d^{\,\ell m}_{\,2}(\io)\,e^{i \,m \,\varphi_0},
\label{ellWF:harm}
\end{equation}
where $d^{\,\ell m}_{\,2}(\io)$ are the Wigner $d$ functions (e.g.,~\cite{Wiaux:2005fm}).
The spherical harmonic modes of the waveform in time-domain have the following generic form
\be
\label{eq:hlmt}
\hlm(t)= A_{\ell m}(t)\,e^{i\,\varphi_{\ell m}(t)}
\ee
Further, $m<0$ modes are related to $m>0$ modes as $\h_{\ell, -m}(t)=(-)^{\ell}\h_{\ell m}^*(t)$.
Using Eq.~\eqref{ellWF:harm} and Eq.~\eqref{eq:hlmt} in Eq.~\eqref{eq:complexh} and making use 
of the above property we can write expressions for the real and imaginary part as 

\begin{widetext}
\begin{subequations}
\label{eq:hphc_timedomain_2}
\begin{align}
h_+(t) & = \sum^{+\infty}_{\ell=2}\sum^{\ell}_{m=1}\sqrt{\frac{2\ell+1}{4\pi}}\, \Bigr[(-)^{\ell}d^{\,\ell ,-m}_{\,2}(\io)+d^{\,\ell m}_{\,2}(\io) \Bigr]\,
A_{\ell m}(t)\,\cos[\varphi_{\ell m}(t)+m \varphi_0],\\
h_{\times}(t) & = \sum^{+\infty}_{\ell=2}\sum^{\ell}_{m=1}
\sqrt{\frac{2\ell+1}{4\pi}}\,\Bigl[(-)^{\ell}d^{\,\ell ,-m}_{\,2}(\io)-d^{\,\ell m}_{\,2}(\io)\Bigr]\,
A_{\ell m}(t)\,\sin[\varphi_{\ell m}(t)+m \varphi_0].
\end{align}
\end{subequations}
\end{widetext}
The frequency domain $+$ and $\times$ waveforms can now be obtained simply by taking Fourier Transform of $h_+(t)$
and $h_{\times}(t)$, respectively
\begin{widetext}
\begin{subequations}
\label{eq:hphc_freqdomain}
\begin{align}
{\tilde h}_+(f) & = \sum^{+\infty}_{\ell=2}\sum^{\ell}_{m=1}
\sqrt{\frac{2\ell+1}{4\pi}}\, \Bigr[(-)^{\ell}d^{\,\ell ,-m}_{\,2}(\io)+d^{\,\ell m}_{\,2}(\io) \Bigr]\,
\biggl\{\cos(m\varphi_0)\,\tilde{h}^\mathrm{R}_{\ell m}(f)-\sin(m\varphi_0)\,\tilde{h}^\mathrm{I}_{\ell m}(f)\biggr\},\\
{\tilde h}_{\times}(f) & = \sum^{+\infty}_{\ell=2}\sum^{\ell}_{m=1}
\sqrt{\frac{2\ell+1}{4\pi}}\,\Bigl[(-)^{\ell}d^{\,\ell ,-m}_{\,2}(\io)-d^{\,\ell m}_{\,2}(\io)\Bigr]\,
\biggl\{\sin(m\varphi_0)\,\tilde{h}^\mathrm{R}_{\ell m}(f)+\cos(m\varphi_0)\,\tilde{h}^\mathrm{I}_{\ell m}(f)\biggr\}.
\end{align}
\end{subequations}
\end{widetext}
where $h^\mathrm{R}_{\ell m}(f)$ and $h^\mathrm{I}_{\ell m}(f)$ are the Fourier transforms of the real and imaginary parts of $\hlm(t)$. 
\begin{subequations}
\label{eq:hlmf}
\begin{align}
h^\mathrm{R}_{\ell m}(f) &=\int^{\infty}_{-\infty} e^{2 \pi i f t} A_{\ell m}(t) \cos \varphi_{\ell m}(t)\,dt,\\ 
h^\mathrm{I}_{\ell m}(f) &=\int^{\infty}_{-\infty} e^{2 \pi i f t} A_{\ell m}(t) \sin \varphi_{\ell m}(t)\,dt. 
\end{align}
\end{subequations}
We know that for non-spinning binaries (as well as for non-precessing binaries), $h^\mathrm{I}_{\ell m}(f) = -i h^\mathrm{R}_{\ell m}(f)$. This allows us to write Eq.~(\ref{eq:hphc_freqdomain}) as 

\begin{subequations}
\label{eq:hphc_freqdomain_final}
\begin{align}
{\tilde h}_+(f) & = \sum^{+\infty}_{\ell=2}\sum^{\ell}_{m=1} \, \Bigr[(-)^\ell\frac{d^{\,\ell ,-m}_{\,2}(\io)}{d^{\,\ell m}_{\,2}(\io)} + 1\Bigr] \, Y^{\ell m}_{-2} (\iota, \varphi_0) \, \tilde{h}^\mathrm{R}_{\ell m}(f) \\ 
{\tilde h}_\times(f) & = -i \sum^{+\infty}_{\ell=2}\sum^{\ell}_{m=1} \, \Bigr[(-)^\ell\frac{d^{\,\ell ,-m}_{\,2}(\io)}{d^{\,\ell m}_{\,2}(\io)} - 1\Bigr] \, Y^{\ell m}_{-2} (\iota, \varphi_0) \, \tilde{h}^\mathrm{R}_{\ell m}(f).
\end{align}
\end{subequations}
Note that $\tilde{h}^\mathrm{R}_{\ell m}(f)$ can be written as 
\begin{equation}
\tilde{h}^\mathrm{R}_{\ell m}(f) = A_{\ell m}(f) \, e^{i \, \Psi_{\ell m} (f)}. 
\end{equation}
The phenomenological model for the frequency domain amplitudes $A_{\ell m}(f)$ and phases $\Psi_{\ell m}(f)$ are obtained by fitting the FFT of hybrids. 

The signal observed at a detector is a linear combination of the two polarizations $h_+$ and $h_\times$. The Fourier transform of the observed signal can be written in terms of the Fourier transform of the two polarizations as 
\begin{equation}
\tilde{h}(f) = F_+(\theta, \phi, \psi) \, \tilde{h}_+(f)  + F_\times(\theta, \phi, \psi) \, \tilde{h}_\times(f),
\end{equation}
where the antenna pattern functions $F_+(\theta, \phi, \psi)$ and $F_\times(\theta, \phi, \psi)$ are functions of two angles $(\theta, \phi)$ describing the location of the binary in the sky and the polarization angle $\psi$. 

\bibliographystyle{apsrev-nourl}
\bibliography{PhenomHM}

\begin{thebibliography}{59}
\expandafter\ifx\csname natexlab\endcsname\relax\def\natexlab#1{#1}\fi
\expandafter\ifx\csname bibnamefont\endcsname\relax
  \def\bibnamefont#1{#1}\fi
\expandafter\ifx\csname bibfnamefont\endcsname\relax
  \def\bibfnamefont#1{#1}\fi
\expandafter\ifx\csname citenamefont\endcsname\relax
  \def\citenamefont#1{#1}\fi
\expandafter\ifx\csname url\endcsname\relax
  \def\url#1{\texttt{#1}}\fi
\expandafter\ifx\csname urlprefix\endcsname\relax\def\urlprefix{URL }\fi
\providecommand{\bibinfo}[2]{#2}
\providecommand{\eprint}[2][]{\url{#2}}

\bibitem[{\citenamefont{{Abbott et
  al.}}(2016{\natexlab{a}})}]{LSC_2016firstdetection}
\bibinfo{author}{\bibfnamefont{B.~P.} \bibnamefont{{Abbott et al.}}}
  (\bibinfo{collaboration}{LIGO Scientific Collaboration and Virgo
  Collaboration}), \bibinfo{journal}{Phys. Rev. Lett.}
  \textbf{\bibinfo{volume}{116}}, \bibinfo{pages}{061102}
  (\bibinfo{year}{2016}{\natexlab{a}}).

\bibitem[{\citenamefont{{Abbott et
  al.}}(2016{\natexlab{b}})}]{LSC_2016seconddetection}
\bibinfo{author}{\bibfnamefont{B.~P.} \bibnamefont{{Abbott et al.}}}
  (\bibinfo{collaboration}{LIGO Scientific Collaboration and Virgo
  Collaboration}), \bibinfo{journal}{Phys. Rev. Lett.}
  \textbf{\bibinfo{volume}{116}}, \bibinfo{pages}{241103}
  (\bibinfo{year}{2016}{\natexlab{b}}).

\bibitem[{\citenamefont{Abbott
  et~al.}(2017{\natexlab{a}})}]{PhysRevLett.118.221101}
\bibinfo{author}{\bibfnamefont{B.~P.} \bibnamefont{Abbott}}
  \bibnamefont{et~al.} (\bibinfo{collaboration}{LIGO Scientific and Virgo
  Collaboration}), \bibinfo{journal}{Phys. Rev. Lett.}
  \textbf{\bibinfo{volume}{118}}, \bibinfo{pages}{221101}
  (\bibinfo{year}{2017}{\natexlab{a}}).

\bibitem[{\citenamefont{{Abbott et al.\ (LIGO Scientific
  Collaboration)}}(2016)}]{LSC_2016rates}
\bibinfo{author}{\bibfnamefont{B.~P.} \bibnamefont{{Abbott et al.\ (LIGO
  Scientific Collaboration)}}}, \bibinfo{journal}{ArXiv e-prints}
  (\bibinfo{year}{2016}).

\bibitem[{\citenamefont{{Abbott et
  al.}}(2016{\natexlab{c}})}]{LSC_2016O1results}
\bibinfo{author}{\bibfnamefont{B.~P.} \bibnamefont{{Abbott et al.}}}
  (\bibinfo{collaboration}{LIGO Scientific Collaboration and Virgo
  Collaboration}), \bibinfo{journal}{Phys. Rev. X}
  \textbf{\bibinfo{volume}{6}}, \bibinfo{pages}{041015}
  (\bibinfo{year}{2016}{\natexlab{c}}).

\bibitem[{\citenamefont{Usman et~al.}(2016)}]{Usman:2015kfa}
\bibinfo{author}{\bibfnamefont{S.~A.} \bibnamefont{Usman}}
  \bibnamefont{et~al.}, \bibinfo{journal}{Class. Quant. Grav.}
  \textbf{\bibinfo{volume}{33}}, \bibinfo{pages}{215004}
  (\bibinfo{year}{2016}).

\bibitem[{\citenamefont{{Messick} et~al.}(2017)\citenamefont{{Messick},
  {Blackburn}, {Brady}, {Brockill}, {Cannon}, {Cariou}, {Caudill},
  {Chamberlin}, {Creighton}, {Everett} et~al.}}]{2017PhRvD..95d2001M}
\bibinfo{author}{\bibfnamefont{C.}~\bibnamefont{{Messick}}},
  \bibinfo{author}{\bibfnamefont{K.}~\bibnamefont{{Blackburn}}},
  \bibinfo{author}{\bibfnamefont{P.}~\bibnamefont{{Brady}}},
  \bibinfo{author}{\bibfnamefont{P.}~\bibnamefont{{Brockill}}},
  \bibinfo{author}{\bibfnamefont{K.}~\bibnamefont{{Cannon}}},
  \bibinfo{author}{\bibfnamefont{R.}~\bibnamefont{{Cariou}}},
  \bibinfo{author}{\bibfnamefont{S.}~\bibnamefont{{Caudill}}},
  \bibinfo{author}{\bibfnamefont{S.~J.} \bibnamefont{{Chamberlin}}},
  \bibinfo{author}{\bibfnamefont{J.~D.~E.} \bibnamefont{{Creighton}}},
  \bibinfo{author}{\bibfnamefont{R.}~\bibnamefont{{Everett}}},
  \bibnamefont{et~al.}, \bibinfo{journal}{\prd} \textbf{\bibinfo{volume}{95}},
  \bibinfo{eid}{042001} (\bibinfo{year}{2017}).

\bibitem[{\citenamefont{Veitch et~al.}(2015)}]{Veitch:2014wba}
\bibinfo{author}{\bibfnamefont{J.}~\bibnamefont{Veitch}} \bibnamefont{et~al.},
  \bibinfo{journal}{Phys. Rev.} \textbf{\bibinfo{volume}{D91}},
  \bibinfo{pages}{042003} (\bibinfo{year}{2015}).

\bibitem[{\citenamefont{{Abbott et al.}}(2016{\natexlab{d}})}]{LSC_2016grtests}
\bibinfo{author}{\bibfnamefont{B.~P.} \bibnamefont{{Abbott et al.}}}
  (\bibinfo{collaboration}{LIGO Scientific and Virgo Collaborations}),
  \bibinfo{journal}{Phys. Rev. Lett.} \textbf{\bibinfo{volume}{116}},
  \bibinfo{pages}{221101} (\bibinfo{year}{2016}{\natexlab{d}}).

\bibitem[{\citenamefont{Buonanno and Damour}(1999)}]{Buonanno:1998gg}
\bibinfo{author}{\bibfnamefont{A.}~\bibnamefont{Buonanno}} \bibnamefont{and}
  \bibinfo{author}{\bibfnamefont{T.}~\bibnamefont{Damour}},
  \bibinfo{journal}{Phys. Rev.} \textbf{\bibinfo{volume}{D59}},
  \bibinfo{pages}{084006} (\bibinfo{year}{1999}).

\bibitem[{\citenamefont{Buonanno and Damour}(2000)}]{Buonanno:2000ef}
\bibinfo{author}{\bibfnamefont{A.}~\bibnamefont{Buonanno}} \bibnamefont{and}
  \bibinfo{author}{\bibfnamefont{T.}~\bibnamefont{Damour}},
  \bibinfo{journal}{Phys. Rev.} \textbf{\bibinfo{volume}{D62}},
  \bibinfo{pages}{064015} (\bibinfo{year}{2000}).

\bibitem[{\citenamefont{Taracchini et~al.}(2014)}]{Taracchini:2013rva}
\bibinfo{author}{\bibfnamefont{A.}~\bibnamefont{Taracchini}}
  \bibnamefont{et~al.}, \bibinfo{journal}{Phys. Rev.}
  \textbf{\bibinfo{volume}{D89}}, \bibinfo{pages}{061502}
  (\bibinfo{year}{2014}).

\bibitem[{\citenamefont{Bohé et~al.}(2017)}]{Bohe:2016gbl}
\bibinfo{author}{\bibfnamefont{A.}~\bibnamefont{Bohé}} \bibnamefont{et~al.},
  \bibinfo{journal}{Phys. Rev.} \textbf{\bibinfo{volume}{D95}},
  \bibinfo{pages}{044028} (\bibinfo{year}{2017}).

\bibitem[{\citenamefont{Pan et~al.}(2011{\natexlab{a}})\citenamefont{Pan,
  Buonanno, Fujita, Racine, and Tagoshi}}]{Pan:2010hz}
\bibinfo{author}{\bibfnamefont{Y.}~\bibnamefont{Pan}},
  \bibinfo{author}{\bibfnamefont{A.}~\bibnamefont{Buonanno}},
  \bibinfo{author}{\bibfnamefont{R.}~\bibnamefont{Fujita}},
  \bibinfo{author}{\bibfnamefont{E.}~\bibnamefont{Racine}}, \bibnamefont{and}
  \bibinfo{author}{\bibfnamefont{H.}~\bibnamefont{Tagoshi}},
  \bibinfo{journal}{Phys. Rev.} \textbf{\bibinfo{volume}{D83}},
  \bibinfo{pages}{064003} (\bibinfo{year}{2011}{\natexlab{a}}),
  \bibinfo{note}{[Erratum: Phys. Rev.D87,no.10,109901(2013)]}.

\bibitem[{\citenamefont{Pan et~al.}(2011{\natexlab{b}})\citenamefont{Pan,
  Buonanno, Boyle, Buchman, Kidder et~al.}}]{Pan:2011gk}
\bibinfo{author}{\bibfnamefont{Y.}~\bibnamefont{Pan}},
  \bibinfo{author}{\bibfnamefont{A.}~\bibnamefont{Buonanno}},
  \bibinfo{author}{\bibfnamefont{M.}~\bibnamefont{Boyle}},
  \bibinfo{author}{\bibfnamefont{L.~T.} \bibnamefont{Buchman}},
  \bibinfo{author}{\bibfnamefont{L.~E.} \bibnamefont{Kidder}},
  \bibnamefont{et~al.}, \bibinfo{journal}{Phys. Rev. D}
  \textbf{\bibinfo{volume}{84}}, \bibinfo{pages}{124052}
  (\bibinfo{year}{2011}{\natexlab{b}}).

\bibitem[{\citenamefont{Pan et~al.}(2014)\citenamefont{Pan, Buonanno,
  Taracchini, Kidder, Mroué, Pfeiffer, Scheel, and Szilágyi}}]{Pan:2013rra}
\bibinfo{author}{\bibfnamefont{Y.}~\bibnamefont{Pan}},
  \bibinfo{author}{\bibfnamefont{A.}~\bibnamefont{Buonanno}},
  \bibinfo{author}{\bibfnamefont{A.}~\bibnamefont{Taracchini}},
  \bibinfo{author}{\bibfnamefont{L.~E.} \bibnamefont{Kidder}},
  \bibinfo{author}{\bibfnamefont{A.~H.} \bibnamefont{Mroué}},
  \bibinfo{author}{\bibfnamefont{H.~P.} \bibnamefont{Pfeiffer}},
  \bibinfo{author}{\bibfnamefont{M.~A.} \bibnamefont{Scheel}},
  \bibnamefont{and}
  \bibinfo{author}{\bibfnamefont{B.}~\bibnamefont{Szilágyi}},
  \bibinfo{journal}{Phys. Rev.} \textbf{\bibinfo{volume}{D89}},
  \bibinfo{pages}{084006} (\bibinfo{year}{2014}).

\bibitem[{\citenamefont{Damour and Nagar}(2009)}]{Damour:2009kr}
\bibinfo{author}{\bibfnamefont{T.}~\bibnamefont{Damour}} \bibnamefont{and}
  \bibinfo{author}{\bibfnamefont{A.}~\bibnamefont{Nagar}},
  \bibinfo{journal}{Phys. Rev. D} \textbf{\bibinfo{volume}{79}},
  \bibinfo{pages}{081503} (\bibinfo{year}{2009}).

\bibitem[{\citenamefont{Damour et~al.}(2009)\citenamefont{Damour, Iyer, and
  Nagar}}]{Damour:2008gu}
\bibinfo{author}{\bibfnamefont{T.}~\bibnamefont{Damour}},
  \bibinfo{author}{\bibfnamefont{B.~R.} \bibnamefont{Iyer}}, \bibnamefont{and}
  \bibinfo{author}{\bibfnamefont{A.}~\bibnamefont{Nagar}},
  \bibinfo{journal}{Phys. Rev.} \textbf{\bibinfo{volume}{D79}},
  \bibinfo{pages}{064004} (\bibinfo{year}{2009}).

\bibitem[{\citenamefont{Damour et~al.}(2013)\citenamefont{Damour, Nagar, and
  Bernuzzi}}]{Damour:2012ky}
\bibinfo{author}{\bibfnamefont{T.}~\bibnamefont{Damour}},
  \bibinfo{author}{\bibfnamefont{A.}~\bibnamefont{Nagar}}, \bibnamefont{and}
  \bibinfo{author}{\bibfnamefont{S.}~\bibnamefont{Bernuzzi}},
  \bibinfo{journal}{Phys. Rev.} \textbf{\bibinfo{volume}{D87}},
  \bibinfo{pages}{084035} (\bibinfo{year}{2013}).

\bibitem[{\citenamefont{Ajith et~al.}(2008)}]{Ajith:2007kx}
\bibinfo{author}{\bibfnamefont{P.}~\bibnamefont{Ajith}} \bibnamefont{et~al.},
  \bibinfo{journal}{Phys. Rev. D} \textbf{\bibinfo{volume}{77}},
  \bibinfo{pages}{104017} (\bibinfo{year}{2008}).

\bibitem[{\citenamefont{Ajith et~al.}(2011)\citenamefont{Ajith, Hannam, Husa,
  Chen, Br\"ugmann, Dorband, M\"uller, Ohme, Pollney, Reisswig
  et~al.}}]{Ajith:2009bn}
\bibinfo{author}{\bibfnamefont{P.}~\bibnamefont{Ajith}},
  \bibinfo{author}{\bibfnamefont{M.}~\bibnamefont{Hannam}},
  \bibinfo{author}{\bibfnamefont{S.}~\bibnamefont{Husa}},
  \bibinfo{author}{\bibfnamefont{Y.}~\bibnamefont{Chen}},
  \bibinfo{author}{\bibfnamefont{B.}~\bibnamefont{Br\"ugmann}},
  \bibinfo{author}{\bibfnamefont{N.}~\bibnamefont{Dorband}},
  \bibinfo{author}{\bibfnamefont{D.}~\bibnamefont{M\"uller}},
  \bibinfo{author}{\bibfnamefont{F.}~\bibnamefont{Ohme}},
  \bibinfo{author}{\bibfnamefont{D.}~\bibnamefont{Pollney}},
  \bibinfo{author}{\bibfnamefont{C.}~\bibnamefont{Reisswig}},
  \bibnamefont{et~al.}, \bibinfo{journal}{Phys. Rev. Lett.}
  \textbf{\bibinfo{volume}{106}}, \bibinfo{pages}{241101}
  (\bibinfo{year}{2011}).

\bibitem[{\citenamefont{Santamaria et~al.}(2010)}]{Santamaria:2010yb}
\bibinfo{author}{\bibfnamefont{L.}~\bibnamefont{Santamaria}}
  \bibnamefont{et~al.}, \bibinfo{journal}{Phys. Rev.}
  \textbf{\bibinfo{volume}{D82}}, \bibinfo{pages}{064016}
  (\bibinfo{year}{2010}).

\bibitem[{\citenamefont{Hannam et~al.}(2014)\citenamefont{Hannam, Schmidt,
  Bohé, Haegel, Husa, Ohme, Pratten, and Pürrer}}]{Hannam:2013oca}
\bibinfo{author}{\bibfnamefont{M.}~\bibnamefont{Hannam}},
  \bibinfo{author}{\bibfnamefont{P.}~\bibnamefont{Schmidt}},
  \bibinfo{author}{\bibfnamefont{A.}~\bibnamefont{Bohé}},
  \bibinfo{author}{\bibfnamefont{L.}~\bibnamefont{Haegel}},
  \bibinfo{author}{\bibfnamefont{S.}~\bibnamefont{Husa}},
  \bibinfo{author}{\bibfnamefont{F.}~\bibnamefont{Ohme}},
  \bibinfo{author}{\bibfnamefont{G.}~\bibnamefont{Pratten}}, \bibnamefont{and}
  \bibinfo{author}{\bibfnamefont{M.}~\bibnamefont{Pürrer}},
  \bibinfo{journal}{Phys. Rev. Lett.} \textbf{\bibinfo{volume}{113}},
  \bibinfo{pages}{151101} (\bibinfo{year}{2014}).

\bibitem[{\citenamefont{Khan et~al.}(2016)\citenamefont{Khan, Husa, Hannam,
  Ohme, P\"urrer, Forteza, and Boh\'e}}]{Khan_2016IMRPhenomD}
\bibinfo{author}{\bibfnamefont{S.}~\bibnamefont{Khan}},
  \bibinfo{author}{\bibfnamefont{S.}~\bibnamefont{Husa}},
  \bibinfo{author}{\bibfnamefont{M.}~\bibnamefont{Hannam}},
  \bibinfo{author}{\bibfnamefont{F.}~\bibnamefont{Ohme}},
  \bibinfo{author}{\bibfnamefont{M.}~\bibnamefont{P\"urrer}},
  \bibinfo{author}{\bibfnamefont{X.~J.} \bibnamefont{Forteza}},
  \bibnamefont{and} \bibinfo{author}{\bibfnamefont{A.}~\bibnamefont{Boh\'e}},
  \bibinfo{journal}{Phys. Rev. D} \textbf{\bibinfo{volume}{93}},
  \bibinfo{pages}{044007} (\bibinfo{year}{2016}).

\bibitem[{\citenamefont{Husa et~al.}(2016)\citenamefont{Husa, Khan, Hannam,
  P\"urrer, Ohme, Forteza, and Boh\'e}}]{Husa_2016IMRPhenomD}
\bibinfo{author}{\bibfnamefont{S.}~\bibnamefont{Husa}},
  \bibinfo{author}{\bibfnamefont{S.}~\bibnamefont{Khan}},
  \bibinfo{author}{\bibfnamefont{M.}~\bibnamefont{Hannam}},
  \bibinfo{author}{\bibfnamefont{M.}~\bibnamefont{P\"urrer}},
  \bibinfo{author}{\bibfnamefont{F.}~\bibnamefont{Ohme}},
  \bibinfo{author}{\bibfnamefont{X.~J.} \bibnamefont{Forteza}},
  \bibnamefont{and} \bibinfo{author}{\bibfnamefont{A.}~\bibnamefont{Boh\'e}},
  \bibinfo{journal}{Phys. Rev. D} \textbf{\bibinfo{volume}{93}},
  \bibinfo{pages}{044006} (\bibinfo{year}{2016}).

\bibitem[{\citenamefont{Abbott et~al.}(2017{\natexlab{b}})}]{Abbott:2016wiq}
\bibinfo{author}{\bibfnamefont{B.~P.} \bibnamefont{Abbott}}
  \bibnamefont{et~al.} (\bibinfo{collaboration}{Virgo, LIGO Scientific}),
  \bibinfo{journal}{Class. Quant. Grav.} \textbf{\bibinfo{volume}{34}},
  \bibinfo{pages}{104002} (\bibinfo{year}{2017}{\natexlab{b}}).

\bibitem[{\citenamefont{Varma et~al.}(2014)\citenamefont{Varma, Ajith, Husa,
  Bustillo, Hannam, and P\"urrer}}]{Varma:2014hm}
\bibinfo{author}{\bibfnamefont{V.}~\bibnamefont{Varma}},
  \bibinfo{author}{\bibfnamefont{P.}~\bibnamefont{Ajith}},
  \bibinfo{author}{\bibfnamefont{S.}~\bibnamefont{Husa}},
  \bibinfo{author}{\bibfnamefont{J.~C.} \bibnamefont{Bustillo}},
  \bibinfo{author}{\bibfnamefont{M.}~\bibnamefont{Hannam}}, \bibnamefont{and}
  \bibinfo{author}{\bibfnamefont{M.}~\bibnamefont{P\"urrer}},
  \bibinfo{journal}{Phys. Rev. D} \textbf{\bibinfo{volume}{90}},
  \bibinfo{pages}{124004} (\bibinfo{year}{2014}).

\bibitem[{\citenamefont{Calder\'on~Bustillo
  et~al.}(2016)\citenamefont{Calder\'on~Bustillo, Husa, Sintes, and
  P\"urrer}}]{CalderonBustillo:2016hm}
\bibinfo{author}{\bibfnamefont{J.}~\bibnamefont{Calder\'on~Bustillo}},
  \bibinfo{author}{\bibfnamefont{S.}~\bibnamefont{Husa}},
  \bibinfo{author}{\bibfnamefont{A.~M.} \bibnamefont{Sintes}},
  \bibnamefont{and} \bibinfo{author}{\bibfnamefont{M.}~\bibnamefont{P\"urrer}},
  \bibinfo{journal}{Phys. Rev. D} \textbf{\bibinfo{volume}{93}},
  \bibinfo{pages}{084019} (\bibinfo{year}{2016}).

\bibitem[{\citenamefont{Varma and Ajith}(2016)}]{Varma:2016dnf}
\bibinfo{author}{\bibfnamefont{V.}~\bibnamefont{Varma}} \bibnamefont{and}
  \bibinfo{author}{\bibfnamefont{P.}~\bibnamefont{Ajith}}
  (\bibinfo{year}{2016}).

\bibitem[{\citenamefont{Sintes and Vecchio}(2000)}]{Sintes:1999cg}
\bibinfo{author}{\bibfnamefont{A.~M.} \bibnamefont{Sintes}} \bibnamefont{and}
  \bibinfo{author}{\bibfnamefont{A.}~\bibnamefont{Vecchio}}, in
  \emph{\bibinfo{booktitle}{{Proceedings, 34th Rencontres de Moriond
  gravitational waves and experimental gravity: Les Arcs, France, Jan 23-30,
  1999}}} (\bibinfo{year}{2000}), pp. \bibinfo{pages}{73--78},
  \eprint{gr-qc/0005058},
  \urlprefix\url{http://alice.cern.ch/format/showfull?sysnb=2187848}.

\bibitem[{\citenamefont{Van Den~Broeck and
  Sengupta}(2007{\natexlab{a}})}]{VanDenBroeck:2006ar}
\bibinfo{author}{\bibfnamefont{C.}~\bibnamefont{Van Den~Broeck}}
  \bibnamefont{and} \bibinfo{author}{\bibfnamefont{A.~S.}
  \bibnamefont{Sengupta}}, \bibinfo{journal}{Class. Quant. Grav.}
  \textbf{\bibinfo{volume}{24}}, \bibinfo{pages}{1089}
  (\bibinfo{year}{2007}{\natexlab{a}}).

\bibitem[{\citenamefont{Arun et~al.}(2007)\citenamefont{Arun, Iyer,
  Sathyaprakash, Sinha, and Broeck}}]{Arun:2007hu}
\bibinfo{author}{\bibfnamefont{K.}~\bibnamefont{Arun}},
  \bibinfo{author}{\bibfnamefont{B.~R.} \bibnamefont{Iyer}},
  \bibinfo{author}{\bibfnamefont{B.}~\bibnamefont{Sathyaprakash}},
  \bibinfo{author}{\bibfnamefont{S.}~\bibnamefont{Sinha}}, \bibnamefont{and}
  \bibinfo{author}{\bibfnamefont{C.~V.~D.} \bibnamefont{Broeck}},
  \bibinfo{journal}{Phys.Rev.} \textbf{\bibinfo{volume}{D76}},
  \bibinfo{pages}{104016} (\bibinfo{year}{2007}).

\bibitem[{\citenamefont{Trias and Sintes}(2008)}]{Trias:2008pu}
\bibinfo{author}{\bibfnamefont{M.}~\bibnamefont{Trias}} \bibnamefont{and}
  \bibinfo{author}{\bibfnamefont{A.~M.} \bibnamefont{Sintes}},
  \bibinfo{journal}{Class. Quant. Grav.} \textbf{\bibinfo{volume}{25}},
  \bibinfo{pages}{184032} (\bibinfo{year}{2008}).

\bibitem[{\citenamefont{Arun et~al.}(2009)}]{Arun:2008zn}
\bibinfo{author}{\bibfnamefont{K.~G.} \bibnamefont{Arun}} \bibnamefont{et~al.},
  \bibinfo{journal}{Class. Quant. Grav.} \textbf{\bibinfo{volume}{26}},
  \bibinfo{pages}{094027} (\bibinfo{year}{2009}).

\bibitem[{\citenamefont{Graff et~al.}(2015)\citenamefont{Graff, Buonanno, and
  Sathyaprakash}}]{Graff:2015bba}
\bibinfo{author}{\bibfnamefont{P.~B.} \bibnamefont{Graff}},
  \bibinfo{author}{\bibfnamefont{A.}~\bibnamefont{Buonanno}}, \bibnamefont{and}
  \bibinfo{author}{\bibfnamefont{B.~S.} \bibnamefont{Sathyaprakash}},
  \bibinfo{journal}{Phys. Rev.} \textbf{\bibinfo{volume}{D92}},
  \bibinfo{pages}{022002} (\bibinfo{year}{2015}).

\bibitem[{\citenamefont{Lange et~al.}(2017)}]{Lange:2017wki}
\bibinfo{author}{\bibfnamefont{J.}~\bibnamefont{Lange}} \bibnamefont{et~al.}
  (\bibinfo{year}{2017}).

\bibitem[{\citenamefont{O'Shaughnessy et~al.}(2017)\citenamefont{O'Shaughnessy,
  Blackman, and Field}}]{OShaughnessy:2017tak}
\bibinfo{author}{\bibfnamefont{R.}~\bibnamefont{O'Shaughnessy}},
  \bibinfo{author}{\bibfnamefont{J.}~\bibnamefont{Blackman}}, \bibnamefont{and}
  \bibinfo{author}{\bibfnamefont{S.~E.} \bibnamefont{Field}}
  (\bibinfo{year}{2017}).

\bibitem[{\citenamefont{Mishra et~al.}(2010)\citenamefont{Mishra, Arun, Iyer,
  and Sathyaprakash}}]{Mishra:2010tp}
\bibinfo{author}{\bibfnamefont{C.~K.} \bibnamefont{Mishra}},
  \bibinfo{author}{\bibfnamefont{K.~G.} \bibnamefont{Arun}},
  \bibinfo{author}{\bibfnamefont{B.~R.} \bibnamefont{Iyer}}, \bibnamefont{and}
  \bibinfo{author}{\bibfnamefont{B.~S.} \bibnamefont{Sathyaprakash}},
  \bibinfo{journal}{Phys. Rev.} \textbf{\bibinfo{volume}{D82}},
  \bibinfo{pages}{064010} (\bibinfo{year}{2010}).

\bibitem[{\citenamefont{Krishnendu et~al.}(2017)\citenamefont{Krishnendu, Arun,
  and Mishra}}]{Krishnendu:2017shb}
\bibinfo{author}{\bibfnamefont{N.~V.} \bibnamefont{Krishnendu}},
  \bibinfo{author}{\bibfnamefont{K.~G.} \bibnamefont{Arun}}, \bibnamefont{and}
  \bibinfo{author}{\bibfnamefont{C.~K.} \bibnamefont{Mishra}}
  (\bibinfo{year}{2017}).

\bibitem[{\citenamefont{Ajith et~al.}(2007)}]{Ajith:2007qp}
\bibinfo{author}{\bibfnamefont{P.}~\bibnamefont{Ajith}} \bibnamefont{et~al.},
  \bibinfo{journal}{Class. Quant. Grav.} \textbf{\bibinfo{volume}{24}},
  \bibinfo{pages}{S689} (\bibinfo{year}{2007}).

\bibitem[{\citenamefont{Ajith}(2008)}]{Ajith:2007xh}
\bibinfo{author}{\bibfnamefont{P.}~\bibnamefont{Ajith}},
  \bibinfo{journal}{Class. Quant. Grav.} \textbf{\bibinfo{volume}{25}},
  \bibinfo{pages}{114033} (\bibinfo{year}{2008}).

\bibitem[{\citenamefont{London et~al.}(2017)}]{London:2017aa}
\bibinfo{author}{\bibfnamefont{L.}~\bibnamefont{London}} \bibnamefont{et~al.}
  (\bibinfo{year}{2017}), \bibinfo{note}{{LIGO-P1700203-v2}}.

\bibitem[{\citenamefont{Favata}(2010)}]{Favata:2010zu}
\bibinfo{author}{\bibfnamefont{M.}~\bibnamefont{Favata}},
  \bibinfo{journal}{Class. Quant. Grav.} \textbf{\bibinfo{volume}{27}},
  \bibinfo{pages}{084036} (\bibinfo{year}{2010}).

\bibitem[{\citenamefont{Pollney and Reisswig}(2011)}]{Pollney:2010hs}
\bibinfo{author}{\bibfnamefont{D.}~\bibnamefont{Pollney}} \bibnamefont{and}
  \bibinfo{author}{\bibfnamefont{C.}~\bibnamefont{Reisswig}},
  \bibinfo{journal}{Astrophys. J.} \textbf{\bibinfo{volume}{732}},
  \bibinfo{pages}{L13} (\bibinfo{year}{2011}).

\bibitem[{\citenamefont{Faye et~al.}(2012)\citenamefont{Faye, Marsat, Blanchet,
  and Iyer}}]{Faye:2012we}
\bibinfo{author}{\bibfnamefont{G.}~\bibnamefont{Faye}},
  \bibinfo{author}{\bibfnamefont{S.}~\bibnamefont{Marsat}},
  \bibinfo{author}{\bibfnamefont{L.}~\bibnamefont{Blanchet}}, \bibnamefont{and}
  \bibinfo{author}{\bibfnamefont{B.~R.} \bibnamefont{Iyer}},
  \bibinfo{journal}{Class. Quant. Grav.} \textbf{\bibinfo{volume}{29}},
  \bibinfo{pages}{175004} (\bibinfo{year}{2012}).

\bibitem[{\citenamefont{Blanchet et~al.}(2008)\citenamefont{Blanchet, Faye,
  Iyer, and Sinha}}]{Blanchet:2008je}
\bibinfo{author}{\bibfnamefont{L.}~\bibnamefont{Blanchet}},
  \bibinfo{author}{\bibfnamefont{G.}~\bibnamefont{Faye}},
  \bibinfo{author}{\bibfnamefont{B.~R.} \bibnamefont{Iyer}}, \bibnamefont{and}
  \bibinfo{author}{\bibfnamefont{S.}~\bibnamefont{Sinha}},
  \bibinfo{journal}{Class.Quant.Grav.} \textbf{\bibinfo{volume}{25}},
  \bibinfo{pages}{165003} (\bibinfo{year}{2008}).

\bibitem[{\citenamefont{Kidder}(2008)}]{Kidder:2007rt}
\bibinfo{author}{\bibfnamefont{L.~E.} \bibnamefont{Kidder}},
  \bibinfo{journal}{Phys. Rev.} \textbf{\bibinfo{volume}{D77}},
  \bibinfo{pages}{044016} (\bibinfo{year}{2008}).

\bibitem[{\citenamefont{Arun et~al.}(2004)\citenamefont{Arun, Blanchet, Iyer,
  and Qusailah}}]{Arun:2004ff}
\bibinfo{author}{\bibfnamefont{K.~G.} \bibnamefont{Arun}},
  \bibinfo{author}{\bibfnamefont{L.}~\bibnamefont{Blanchet}},
  \bibinfo{author}{\bibfnamefont{B.~R.} \bibnamefont{Iyer}}, \bibnamefont{and}
  \bibinfo{author}{\bibfnamefont{M.~S.~S.} \bibnamefont{Qusailah}},
  \bibinfo{journal}{Class. Quant. Grav.} \textbf{\bibinfo{volume}{21}},
  \bibinfo{pages}{3771} (\bibinfo{year}{2004}), \bibinfo{note}{[Erratum: Class.
  Quant. Grav.22,3115(2005)]}.

\bibitem[{\citenamefont{Blanchet et~al.}(1996)\citenamefont{Blanchet, Iyer,
  Will, and Wiseman}}]{Blanchet:1996pi}
\bibinfo{author}{\bibfnamefont{L.}~\bibnamefont{Blanchet}},
  \bibinfo{author}{\bibfnamefont{B.~R.} \bibnamefont{Iyer}},
  \bibinfo{author}{\bibfnamefont{C.~M.} \bibnamefont{Will}}, \bibnamefont{and}
  \bibinfo{author}{\bibfnamefont{A.~G.} \bibnamefont{Wiseman}},
  \bibinfo{journal}{Class. Quant. Grav.} \textbf{\bibinfo{volume}{13}},
  \bibinfo{pages}{575} (\bibinfo{year}{1996}).

\bibitem[{\citenamefont{Blanchet et~al.}(2004)\citenamefont{Blanchet, Damour,
  Esposito-Farese, and Iyer}}]{Blanchet:2004ek}
\bibinfo{author}{\bibfnamefont{L.}~\bibnamefont{Blanchet}},
  \bibinfo{author}{\bibfnamefont{T.}~\bibnamefont{Damour}},
  \bibinfo{author}{\bibfnamefont{G.}~\bibnamefont{Esposito-Farese}},
  \bibnamefont{and} \bibinfo{author}{\bibfnamefont{B.~R.} \bibnamefont{Iyer}},
  \bibinfo{journal}{Phys. Rev. Lett.} \textbf{\bibinfo{volume}{93}},
  \bibinfo{pages}{091101} (\bibinfo{year}{2004}).

\bibitem[{\citenamefont{Calderón~Bustillo
  et~al.}(2015)\citenamefont{Calderón~Bustillo, Bohé, Husa, Sintes, Hannam,
  and Pürrer}}]{Bustillo:2015ova}
\bibinfo{author}{\bibfnamefont{J.}~\bibnamefont{Calderón~Bustillo}},
  \bibinfo{author}{\bibfnamefont{A.}~\bibnamefont{Bohé}},
  \bibinfo{author}{\bibfnamefont{S.}~\bibnamefont{Husa}},
  \bibinfo{author}{\bibfnamefont{A.~M.} \bibnamefont{Sintes}},
  \bibinfo{author}{\bibfnamefont{M.}~\bibnamefont{Hannam}}, \bibnamefont{and}
  \bibinfo{author}{\bibfnamefont{M.}~\bibnamefont{Pürrer}}
  (\bibinfo{year}{2015}).

\bibitem[{\citenamefont{Berti et~al.}(2009)\citenamefont{Berti, Cardoso, and
  Starinets}}]{Berti:2009kk}
\bibinfo{author}{\bibfnamefont{E.}~\bibnamefont{Berti}},
  \bibinfo{author}{\bibfnamefont{V.}~\bibnamefont{Cardoso}}, \bibnamefont{and}
  \bibinfo{author}{\bibfnamefont{A.~O.} \bibnamefont{Starinets}},
  \bibinfo{journal}{Class. Quant. Grav.} \textbf{\bibinfo{volume}{26}},
  \bibinfo{pages}{163001} (\bibinfo{year}{2009}).

\bibitem[{\citenamefont{Varma et~al.}(2013)\citenamefont{Varma, Fujita,
  Choudhary, and Iyer}}]{Varma:2013kna}
\bibinfo{author}{\bibfnamefont{V.}~\bibnamefont{Varma}},
  \bibinfo{author}{\bibfnamefont{R.}~\bibnamefont{Fujita}},
  \bibinfo{author}{\bibfnamefont{A.}~\bibnamefont{Choudhary}},
  \bibnamefont{and} \bibinfo{author}{\bibfnamefont{B.~R.} \bibnamefont{Iyer}},
  \bibinfo{journal}{Phys. Rev.} \textbf{\bibinfo{volume}{D88}},
  \bibinfo{pages}{024038} (\bibinfo{year}{2013}).

\bibitem[{adl()}]{adligo-psd}
\emph{\bibinfo{title}{Advanced {LIGO} anticipated sensitivity curves}},
  \bibinfo{note}{{LIGO} Document T0900288-v3},
  \urlprefix\url{https://dcc.ligo.org/LIGO-T0900288/public}.

\bibitem[{\citenamefont{Cotesta et~al.}(2017)}]{Cotesta:2017xxx}
\bibinfo{author}{\bibfnamefont{R.}~\bibnamefont{Cotesta}} \bibnamefont{et~al.}
  (\bibinfo{year}{2017}), \bibinfo{note}{in preparation}.

\bibitem[{\citenamefont{Berti and Klein}(2014)}]{Berti:2014fga}
\bibinfo{author}{\bibfnamefont{E.}~\bibnamefont{Berti}} \bibnamefont{and}
  \bibinfo{author}{\bibfnamefont{A.}~\bibnamefont{Klein}},
  \bibinfo{journal}{Phys. Rev.} \textbf{\bibinfo{volume}{D90}},
  \bibinfo{pages}{064012} (\bibinfo{year}{2014}).

\bibitem[{\citenamefont{Damour et~al.}(1998)\citenamefont{Damour, Iyer, and
  Sathyaprakash}}]{DIS98}
\bibinfo{author}{\bibfnamefont{T.}~\bibnamefont{Damour}},
  \bibinfo{author}{\bibfnamefont{B.~R.} \bibnamefont{Iyer}}, \bibnamefont{and}
  \bibinfo{author}{\bibfnamefont{B.~S.} \bibnamefont{Sathyaprakash}},
  \bibinfo{journal}{Phys. Rev. D} \textbf{\bibinfo{volume}{57}},
  \bibinfo{pages}{885} (\bibinfo{year}{1998}).

\bibitem[{\citenamefont{Van Den~Broeck and
  Sengupta}(2007{\natexlab{b}})}]{VanDenBroeck:2006qu}
\bibinfo{author}{\bibfnamefont{C.}~\bibnamefont{Van Den~Broeck}}
  \bibnamefont{and} \bibinfo{author}{\bibfnamefont{A.~S.}
  \bibnamefont{Sengupta}}, \bibinfo{journal}{Class.Quant.Grav.}
  \textbf{\bibinfo{volume}{24}}, \bibinfo{pages}{155}
  (\bibinfo{year}{2007}{\natexlab{b}}).

\bibitem[{\citenamefont{Wiaux et~al.}(2007)\citenamefont{Wiaux, Jacques, and
  Vandergheynst}}]{Wiaux:2005fm}
\bibinfo{author}{\bibfnamefont{Y.}~\bibnamefont{Wiaux}},
  \bibinfo{author}{\bibfnamefont{L.}~\bibnamefont{Jacques}}, \bibnamefont{and}
  \bibinfo{author}{\bibfnamefont{P.}~\bibnamefont{Vandergheynst}},
  \bibinfo{journal}{J. Comput. Phys.} \textbf{\bibinfo{volume}{226}},
  \bibinfo{pages}{2359} (\bibinfo{year}{2007}).

\end{thebibliography}

\end{document}